\documentclass[reprint,pra,aps]{revtex4-2}

\usepackage[utf8]{inputenc}
\usepackage{graphicx}
\usepackage{verbatim}
\usepackage{dsfont}
\usepackage{epstopdf}
\usepackage{natbib}
\usepackage{xcolor}

\usepackage{amsmath,amsfonts,amssymb,amstext}

\def\ket#1{\mathinner{|{#1}\rangle}}
\newcommand{\braket}[2]{\langle#1|#2\rangle}

\begin{document}

\title[]{Quantum walk based state transfer algorithms on the complete $M$-partite graph}

\author{S. Skoupý and M. \v{S}tefa\v{n}\'{a}k}
\affiliation{Department of Physics, Faculty of Nuclear Sciences and Physical Engineering, Czech Technical University in Prague, B\v rehov\'a 7, 115 19 Praha 1 - Star\'e M\v{e}sto, Czech Republic}

\date{\today}

\begin{abstract}
We investigate coined quantum walk search and state transfer algorithms, focusing on the complete $M$-partite graph with $N$ vertices in each partition. First, it is shown that by adding a loop to each vertex the search algorithm finds the marked vertex with unit probability in the limit of a large graph. Next, we employ the evolution operator of the search with two marked vertices to perform a state transfer between the sender and the receiver. We show that when the sender and the receiver are in different partitions the algorithm succeeds with fidelity approaching unity for a large graph. However, when the sender and the receiver are in the same partition the fidelity does not reach exactly one. To amend this problem we propose a state transfer algorithm with an active switch, whose fidelity can be estimated based on the single vertex search alone.
\end{abstract}

\maketitle

\section{Introduction}

Quantum walks \cite{Aharonov1993} are quantum mechanical analogues of random walks. Their dynamics can be formulated either in discrete time steps \cite{Meyer1996},  {as we consider in the present paper}, or in continuous time \cite{Farhi1998}. Both have found promising applications in quantum information processing \cite{aharonov2001}, notably in quantum spatial search where an unsorted database is represented by a graph. The solution to the search problem corresponds to a marked vertex, where the local dynamics is different from the non-marked vertices. Usually, the initial state of the walk is taken as the equal weight superposition of all basis states. The walk is evolved coherently for $T$ steps, after which we perform a measurement which collapses the state of superposition and the walker is found on a single vertex. On various graphs quantum walk is capable of finding the marked vertex with sufficiently high probability in a number of steps that grows with the square root of the number of vertices  {$n$}, i.e., the complexity is the same as for the abstract Grover search algorithm \cite{grover1997} which is known to be optimal. Initially, the investigation was mostly focused on graphs with some degree of symmetry or regularity. Continuous time quantum walks were shown to be optimal \cite{Childs2004} for a complete graph, hypercube and lattices of dimensions greater than 4. Discrete time quantum walks with coins \cite{Ambainis2001} are also optimal on these graphs \cite{shenvi2003,Ambainis2005,Potocek2009,hein2009:search}, in addition, they are optimal for lattices of dimensions greater than 2 \cite{Ambainis2005}. Scattering quantum walk \cite{Hillery2003,Feldman2004,Feldman2007}, which represents an alternative equivalent formulation of the coined quantum walk \cite{andrade2009,venancio2013}, can perform optimal search e.g., on a star graph or a complete $M$-partite graph \cite{reitzner2009}. However, high symmetry is not required for the optimal performance of the quantum walk search \cite{janmark2014,novo2015,meyer2015}. In fact, it was shown \cite{chakraborty2016} that continuous the time quantum walk is optimal on Erdös-Renyi random graphs as long as the probability of an edge existing between any pair of vertices is greater than  {$\left(\log^\frac{3}{2} n \right)/n$}. However, on scale-free networks \cite{albert2002} the application of quantum walk search appears to be limited since the optimal run-time depends on the centrality of the marked node \cite{osada2020}. More recently \cite{chakraborty2020} several sufficient and necessary conditions for continuous time quantum walk search to be optimal were derived.

For the search algorithm (SA) it is not required that we find the marked vertex with unit probability. As long as the success probability is constant, we can repeat the SA several times depending on our error tolerance to find the marked vertex with high probability without changing the overall complexity of the algorithm. Even if the success probability is of the order of  {$1/\log n$}, as for the discrete time quantum walk on the 2D lattice \cite{Ambainis2005}, we can use amplitude amplification \cite{brassard2002} which increases the run-time of the SA by a factor of  {$\sqrt{\log n}$}. There are several graphs where the quantum walk SA is exactly equivalent to the Grover search, e.g., the star graph \cite{reitzner2009}, which means that the success probability is unity. It is interesting that for the discrete time quantum walks the success probability of the SA can be often increased close to unity by adding loops of appropriate weights at each vertex. This was found originally for the complete graph \cite{Ambainis2005} and the hypercube \cite{Potocek2009}. Later investigations \cite{wong2015,wong2018,rhodes2019,rhodes2020,chiang2020} found that this result is much more generic and the optimal weight of the loop depending on the size of the graph and degree of the vertex was identified. Recently, it was proven that adding loops improves the success probability of the SA on all regular locally arc-transitive graphs \cite{hoyer2020}. 

Quantum walks were also applied to the task of state transfer \cite{bose2003} between two vertices of a graph. In this context the initial state of the walk is localized on the sender vertex and we want to transfer it with high probability to the receiver vertex. Provided that the location of the sender and the receiver vertices are known, we can globally design the dynamics such that the walker is transferred from one to the other. This approach was investigated on different graphs such as circle \cite{kurzynski2011,yalcinkaya2015}, 2D lattice \cite{zhan2014}, regular graphs \cite{shang2018} or more general networks \cite{chen2019}. When the sender and the receiver don't know each other's position they can perform state transfer by modifying the local coins at their own vertices, i.e., by implementing the evolution operator of the SA for two marked vertices. This approach was proposed for state transfer on lattices \cite{hein2009} and further analyzed on various types of finite graphs, e. g. on cycles and their variants \cite{kendon2011,barr2014}, star and complete graph with loops \cite{stefanak2016}, complete bipartite graph \cite{stefanak2017}, circulant graphs \cite{zhan2019} or butterfly network \cite{cao2019}. Similar approach can be also applied for finding a path in a maze formed by star graphs \cite{reitzner2017} or trees \cite{koch2018}. 

We consider the state transfer algorithm (STA) following the second approach, i.e., based on the evolution operator of the SA with two marked vertices. For the STA it is desirable that we succeed in the first attempt, i.e., the fidelity of state transfer should be ideally one. Natural candidates for graphs where STA works with unit fidelity are those where also the SA succeeds with certainty. Indeed, the graphs considered in \cite{stefanak2016,stefanak2017} were chosen exactly based on this idea. However, in this paper we show an example of a graph where the SA works with certainty yet in some instance the STA does not have unit fidelity. 

We investigate search and state transfer on the complete $M$-partite graph with $N$ vertices in each partition,  {i.e., the graph has $n=N M$ vertices}. Search on the complete $M$-partite graph was already investigated in \cite{reitzner2009} in the framework of the scattering quantum walk. In the coined walk the success probability of SA reaches $\frac{1}{2}$. We show that by adding a loop to each vertex the success probability tends to one for a large graph. Our approach is based on dimensional reduction \cite{krovi2007,Feldman2010,hillery2012,novo2015}. First, we find an exact invariant subspace $\cal I$ where the state of the algorithm evolves. Next, we investigate the eigenvectors of the evolution operator in the limit of a large graph $M\to \infty$ and $N\to \infty$ and determine those which have non-vanishing overlap with the initial state. These eigenvectors form an orthonormal basis of the relevant part of the invariant subspace. For the SA, we find an exact invariant subspace with dimension 8, and in the asymptotic limit the relevant part is three-dimensional. We then investigate the evolution operator of the search with two marked vertices for the sake of state transfer. There are two possible configurations - either sender and receiver are in the same partition or not. In the first case, we find an exact invariant subspace with dimension 11, while in the second case it has dimension 22. To simplify the calculations, we employ the symmetry of the graph which allows us to exchange the sender and the receiver vertex, or the whole partitions containing them in the latter case. This symmetry splits the invariant subspace $\cal I$ further into two closed subspaces - $\cal I_+$ in which the search with two marked vertices evolves, and the complementary subspace $\cal I_-$ needed for the state transfer. In the configuration where the sender and the receiver are in the same partition, the subspace $\cal I_+$ has dimension 8 and the complementary subspace $\cal I_-$ is three-dimensional. For the second configuration the subspaces have dimensions 12 and 10, respectively. Nevertheless, in the limit of a large graph only five eigenvectors of the evolution operator remain relevant in both configurations - three in the subspace $\cal I_+$ and two in $\cal I_-$. The corresponding eigenvalues can be also determined analytically. We show that when the sender and the receiver are in different partitions the phases of the relevant eigenvalues are harmonic. Hence, state transfer is achieved with unit fidelity. However, when the sender and the receiver are in the same partition, the phases are not harmonic, and the fidelity of state transfer is less than one. To fix this issue we propose an STA with an active switch, where initially only the sender vertex is marked, and after some number of steps the marking is switched to the receiver vertex. The fidelity reachable by this STA can be estimated based on the properties of the search for a single vertex. We show that STA with an active switch achieves perfect state transfer on the complete $M$-partite graph  {in the limit of large $N$ and $M$} for both configurations of the sender and the receiver vertex, and discuss its applicability on other graphs.

The paper is organized as follows: In Section~\ref{sec:2} we describe discrete time quantum walks with coin on finite graphs and introduce the quantum walk search and state transfer algorithms. In Section~\ref{sec:3} search on the complete $M$-partite graph with one marked vertex is investigated in detail. Section~\ref{sec:4} is devoted to the STA. The cases of sender and receiver vertices being in the same or different partitions are analyzed in Section~\ref{sec:4a} and Section~\ref{sec:4b}, respectively. In Section~\ref{sec:5} we consider a STA with an active switch. We conclude and provide an outlook in Section~\ref{sec:6}.

\section{Preliminaries}
\label{sec:2}

In this section we overview the general design of the search and the state transfer algorithms based on the discrete-time quantum walks with coins \cite{shenvi2003}, \cite{Ambainis2005} and \cite{hein2009}. Before we turn to the algorithms, we describe the discrete time quantum walks with coins. Let us start with Hilbert space of the walk. Having a graph $G=\left(V,E\right)$ the corresponding Hilbert space ${\cal H}_G$ can be decomposed as a direct sum
$$
{\cal H}_G = \bigoplus_v {\cal H}_v,
$$
of local Hilbert spaces at each vertex $v\in V$. The orthonormal basis in ${\cal H}_v$ is given by vectors $\ket{v,w}$ such that there is an edge between vertex $v$ and $w$
$$
{\cal H}_v = {\rm Span}\left\lbrace\vert v,w\rangle\vert w\in V ,\left\lbrace v,w\right\rbrace\in E\right\rbrace .
$$
In the basis state $\ket{v,w}$ the first index $v$ describes the actual position of the walker, while the second index $w$ describes the direction of propagation of the walker. Movement of the walker is achieved by application of the flip-flop shift operator $\hat{S}$, which is defined in the following way 
\begin{eqnarray}
\hat{S}\vert v,w\rangle = \vert w,v\rangle 
\label{shift}.
\end{eqnarray}
To generate a nontrivial evolution a coin operator $\hat C$ is applied at every step before the shift takes place. The coin operator can be decomposed into a direct sum
$$
\hat C = \bigoplus_v \hat C_v^{(l)},
$$
where $\hat C_v^{(l)}$ acts locally at a vertex $v$, i.e., it is a unitary operator on ${\cal H}_v$. The evolution operator $\hat{U}$ of one step of the walk is then given by
\begin{eqnarray}
\hat{U}=\hat{S}\hat{C}
\label{evolutOp}.
\end{eqnarray}

The main idea of search and state transfer algorithms is to apply one local coin operator at the marked vertices and different local coin operator at the other vertices of the graph. Marked vertices are those that we want to find or between which we want transfer the walker. Usually, the local coin that is used at non-marked vertices is the Grover operator \cite{grover1997} 
\begin{eqnarray}
\hat{G}_v^{(l)} = 2\vert\Omega_v\rangle\langle\Omega_v\vert -\hat{I}_v^{(l)} ,
\label{grover}
\end{eqnarray}
where $\hat{I}_v^{(l)}$ is an identity operator at the subspace ${\cal H}_v$ and $\vert\Omega_v\rangle$ is an equal weight superposition of all basis states at the vertex $v$ given by
\begin{eqnarray}
\vert\Omega_v\rangle=\frac{1}{\sqrt{d\left(v\right)}}\sum_{\substack{w\\ \left\lbrace v,w\right\rbrace\in E}}\vert v,w\rangle
\label{localSuperposition}.
\end{eqnarray}
Here $d(v)$ denotes the degree of the vertex $v$, which is also the dimension of the subspace ${\cal H}_v$. The coin operator of the SA with one marked vertex $m$ reads
\begin{eqnarray}
\hat{C}_m = \bigoplus_{\substack{v \in V \\ v \neq m}} \hat{G}_v^{(l)} \oplus \hat{C}_m^{(l)}
\label{generalCoinSearch}
\end{eqnarray}
where $\hat{C}_m^{(l)}$ is the local coin operator at the marked vertex. A usual choice for the marked coin is either a simple phase shift by $\pi$ (i.e., $\hat{C}_m^{(l)} = -\hat I_m^{(l)}$), or the Grover operator followed by a phase shift by $\pi$ (i.e., $\hat{C}_m^{(l)} = -\hat G_m^{(l)}$). In the present paper we  consider the latter case. Using the coin operator (\ref{generalCoinSearch}) we obtain the evolution operator of the SA 
$$
\hat{U}_m=\hat{S}\hat{C}_m .
$$
The steps of the SA are: 
\begin{enumerate}
\item Initialize the walk in the equal weight superposition of all basis states 
\begin{eqnarray}
\vert \Omega\rangle = \frac{1}{\sqrt{\sum\limits_{v\in V}
d(v)}}\sum_{v\in V} \sqrt{d(v)}\vert\Omega_v\rangle
\label{InitSearch}.
\end{eqnarray}
\item Apply the evolution operator $\hat{U}_m$ $t$-times. The state of the walk after $t$ steps is given by
$$
\vert\phi\left(t\right)\rangle = \hat U_m^t \ket{\Omega} .
$$
\item Measure the walk.
\end{enumerate}
The probability to find the walker at the marked vertex is given by the summation over all basis states in the subspace ${\cal H}_m$
\begin{eqnarray}
P_m(t) = \sum_{\substack{w\\ \left\lbrace m,w\right\rbrace\in E}}\vert\langle m,w\vert\phi\left(t\right)\rangle\vert^2 .
\label{searchProbab}
\end{eqnarray}
The optimal number of steps $T$ providing high success probability depends on the structure of the graph.

In the case of STA  {we consider $2$ parties, sender and receiver sitting at vertices $s$ and $r$, respectively, which want to establish communication between each other. The sender and the receiver have access only to their local Hilbert spaces ${\cal H}_s$ and ${\cal H}_r$, respectively.} Typical STA \cite{hein2009,stefanak2016,stefanak2017} uses the evolution operator of the search for two marked vertices
$$
\hat U_{s,r} = S \hat C_{s,r},
$$
where we apply the same local coins at both marked vertices at the same time
\begin{eqnarray}
\hat{C}_{s,r} = \bigoplus_{\substack{v \in V \\ v \neq s,r}} \hat G_v^{(l)} \oplus \hat{C}_s^{(l)} \oplus \hat{C}_r^{(l)} 
\label{generalCoinTransfer}.
\end{eqnarray}
However, the initial state of STA is different --- it starts localized at the sender vertex in some state $\ket{s}$. Standard choice is the equal weight superposition of all basis states at the vertex $s$, i.e., $\ket{s} = \ket{\Omega_s}$. The steps of the STA are the following:
\begin{enumerate}
\item Sender initializes the walk at its vertex in the state $\ket{s}$.
\item The evolution operator $\hat{U}_{s,r}$ is applied $t$-times.  The state of the walk after $t$ steps is given by
$$
\vert\phi\left(t\right)\rangle = \hat U_{s,r}^t \ket{s} .
$$
\item Receiver measures the walk at its vertex.
\end{enumerate}
The fidelity of the STA, i.e., the probability that the receiver finds the walker at its vertex, is given by
\begin{eqnarray}
{\cal F}(t) = \sum_{\substack{w\\ \left\lbrace r,w\right\rbrace\in E}}\vert\langle r,w\vert\phi\left(t\right)\rangle\vert^2 .
\label{fidelity}
\end{eqnarray}
The number of steps $T^{(st)}$ required to achieve state transfer with high fidelity depends again on the size and the structure of the graph. In contrast to the SA it is desirable that the STA performs with high fidelity in a single run. In such a case we talk about perfect state transfer.

\section{Search on the complete M-partite graph}
\label{sec:3}

Consider the complete $M$-partite graph (with $M>2$). Complete $M$-partite graph is a graph that has the set of vertices $V$ divided into $M$ subsets, where vertices have no edges between them, but are connected to all vertices in other subsets. We label the vertices of the graph as $v_\alpha$, where $\alpha = 1, \ldots, M$ denotes the partition.  {The basis states of the quantum walk are therefore given by $\ket{v_\alpha,w_\beta}$, $\alpha\neq \beta$.} We also limit ourselves to the case where all parts have the same size $N$, thus the whole graph has $n=MN$ vertices.  {This choice greatly simplifies the construction of the invariant subspace.} The graph is $d$-regular with the vertex degree 
$$
d = N(M-1).
$$
Without loss of generality we assume that the marked vertex is in the first partition.

Search on the complete $M$-partite graph was investigated in \cite{reitzner2009} in the framework of the scattering quantum walk. The difference between the two formulations is that in the coined walk the walker lives on the vertices, while in the scattering walk it lives on the edges. The results of \cite{reitzner2009} can be adopted for the coined quantum walk with a single modification. Namely, for the evaluation of the success probability (\ref{searchProbab}) we consider only the overlap with the states where the walker is at the marked vertex, i.e., of the form  {$\ket{m_1,k_\alpha}$, $k = 1,\ldots, N$, $\alpha = 2,\ldots, M$, which form the basis of the local Hilbert space at marked vertex ${\cal H}_m$. The equal weight superposition of such states corresponds to the state $\ket{w_2}$ in \cite{reitzner2009}.} This gives us the success probability of $\frac{1}{2}$ for a large graph. Note that with probability close to $\frac{1}{2}$ we would find the walker in some state  {$\ket{k_\alpha,m_1}$, $k = 1,\ldots, N$}, $\alpha = 2,\ldots, M$, i.e., where the walker is at the vertex  {$k_\alpha$} and would move to the marked vertex after the application of the shift $\hat S$. In the scattering walk framework these states are also considered in the evaluation of the success probability.  {Nevertheless, we focus on the coined formulation. In the end our goal is to investigate the state transfer between vertices where the coined formulation is more natural, since sender and receiver are considered to be restricted to their local Hilbert spaces ${\cal H}_s$ and ${\cal H}_r$, respectively.} 

We show that one can improve the success probability of the coined walk search by adding one loop at each vertex.  {This is done by adding basis states $\ket{v_\alpha,v_\alpha}$ corresponding to the loop at each vertex, i.e. the dimension of the local Hilbert spaces ${\cal H}_{v_\alpha}$ increases by one. Moreover, we modify the local coin operator (\ref{grover}) by replacing the state $\ket{\Omega_{v_\alpha}}$ (\ref{localSuperposition}) with the state $\ket{\Omega_{v_\alpha}(l)}$ given by 
$$
\ket{\Omega_{v_\alpha}(l)}=\frac{1}{\sqrt{d+l}}\left(\sum_{\substack{\beta = 1 \\ \beta\neq\alpha}}^M\sum_{k=1}^N\ket{ {v_\alpha},k_\beta}+\sqrt{l}\ket{{v_\alpha},{v_\alpha}}\right),
$$
where $l$ is the weight of the loop. Note, however, that the initial state of the SA remains the same as before, i.e., we prepare the system in the equal weight superposition (\ref{InitSearch}) of all basis states excluding the states corresponding to the loops. 
}According to \cite{rhodes2020,hoyer2020} the optimal weight $l$ of the loops is given by
\begin{equation}
\label{weight}
l = \frac{d}{NM} =  {1 - \frac{1}{M}}.     
\end{equation}
Since we focus on the limit of a large graph we put $l=1$. This choice corresponds to the local coin operator being the Grover operator (\ref{grover}) of dimensions $d+1$, except for the marked vertex where we include an additional phase shift by $\pi$. First, we show that the SA evolves in an 8-dimensional invariant subspace ${\cal I}$. We note that in principle the dimension of the exact invariant subspace can be reduced further, since the evolution operator restricted on $\cal I$ still has some degenerate eigenvalues. However, the construction of the basis would not be as intuitive and the ensuing calculations will not simplify. Second, we consider a limit of a large graph which effectively reduces the dimension of the invariant subspace to three. 

Let us begin with the construction of the basis of the invariant subspace ${\cal I}$. The numerical simulations indicate that the state of the walk $\ket{\phi(t)}$ evolves periodically close to a state corresponding to the loop on the marked vertex $\ket{m_1,m_1}$. Hence, we consider this desired target state of the SA as the first basis vector of the invariant subspace
\begin{equation}
\label{search:nu1}
\ket{\nu_1}  = \ket{m_1,m_1}.    
\end{equation}
Next, we add an equal weight superposition of all edges leaving the marked vertex
\begin{equation}
\label{search:nu2}
\vert\nu_2\rangle = \frac{1}{\sqrt{d}}\sum_{\alpha=2}^M\sum_{k=1}^N\vert m_1,k_\alpha\rangle .    
\end{equation}
Concerning the non-marked vertices in the first partition, we add two states corresponding to a superposition of all loops and a superposition of all edges leaving the first partition
\begin{eqnarray}
\nonumber \vert\nu_3\rangle & = & \frac{1}{\sqrt{(N-1)}}\sum_{j \neq m}^N\vert j_1,j_1\rangle , \\
\nonumber \vert\nu_4\rangle & = & \frac{1}{\sqrt{d(N-1)}}\sum_{j \neq m}^N\sum_{\alpha=2}^M\sum_{k=1}^N\vert j_1,k_\alpha\rangle .
\end{eqnarray}
Next, we consider the edges leading to the first partition from the outside and ending either on the marked or non-marked vertex, and construct the following two basis states 
\begin{eqnarray}
\label{search:nu5} \vert\nu_5\rangle & = & \frac{1}{\sqrt{d}}\sum_{\alpha=2}^M\sum_{k=1}^N\vert k_\alpha,m_1\rangle , \\
\nonumber \vert\nu_6\rangle & = & 
\frac{1}{\sqrt{d(N-1)}}\sum_{j \neq m}^N\sum_{\alpha=2}^M\sum_{k=1}^N\vert k_\alpha,j_1\rangle .
\end{eqnarray}
These states can be obtained by applying the shift operator on $\ket{\nu_2}$ and $\ket{\nu_4}$. To complete the basis we consider the states corresponding to the superposition of all edges between the vertices outside of the first partition
$$
\vert\nu_7\rangle = \frac{1}{\sqrt{d(d-N)}}\sum_{\substack{\alpha,\beta=2\\ \beta\neq\alpha}}^M\sum_{j,k=1}^N\vert j_\alpha,k_\beta\rangle
$$
and the superposition of all remaining loops
$$
\vert\nu_8\rangle = \frac{1}{\sqrt{d}}\sum_{\alpha=2}^M\sum_{k=1}^N\vert k_\alpha,k_\alpha\rangle .
$$

Clearly, the initial state of the SA (\ref{InitSearch}) lies in ${\cal I}$ and has the following form
 {
\begin{eqnarray}
\nonumber \ket{\Omega} & = & \frac{1}{\sqrt{MN}}\left[ \vert\nu_2\rangle + \vert\nu_5\rangle +  \sqrt{d-N}\vert\nu_7\rangle + \right.\\
 & & \left. + \sqrt{N-1} (\ket{\nu_4}  + \ket{\nu_6}) \right]
\label{initInvariant}.
\end{eqnarray}}
By direct calculation one can show that the evolution operator $\hat U_m$ of the SA  acts on the basis states according to
\begin{eqnarray}
\nonumber \hat U_m\ket{\nu_1} & = & \frac{1}{d+1}\left((d-1)\ket{\nu_1} - 2\sqrt{d} \ket{\nu_5}\right), \\
\nonumber \hat U_m\ket{\nu_2} & = & -\frac{1}{d+1}\left(2\sqrt{d} \ket{\nu_1} + (d-1) \ket{\nu_5}\right), \\
\nonumber \hat U_m\ket{\nu_3} & = & \frac{1}{d+1}\left((1-d)\ket{\nu_3} +  2\sqrt{d}\ket{\nu_6}\right), \\
\nonumber \hat U_m\ket{\nu_4} & = & \frac{1}{d+1}\left(2\sqrt{d}\ket{\nu_3} +  (d - 1)\ket{\nu_6}\right), \\
\nonumber \hat U_m\ket{\nu_5} & = & \frac{1}{d+1}\left((1-d) \ket{\nu_2} + 2\sqrt{N-1} \ket{\nu_4} + \right.\\
\nonumber & & \left. + 2\sqrt{d-N} \ket{\nu_7} + 2 \ket{\nu_8}\right) , \\
\nonumber \hat U_m\ket{\nu_6} & = & \frac{1}{d+1}\left(2\sqrt{N-1} \ket{\nu_2} - (d+3-2N) \ket{\nu_4} + \right. \\
\nonumber & &  + 2\sqrt{(d-N)(N-1)}\ket{\nu_7} + \\
\nonumber & & \left. + 2\sqrt{N-1}\ket{\nu_8}\right), \\
\nonumber \hat U_m \ket{\nu_7} & = & \frac{1}{d+1}\left(2\sqrt{d-N} \ket{\nu_2}  +  \right. \\
\nonumber & & \left. + 2\sqrt{(d-N)(N-1)} \ket{\nu_4} + \right. \\
\nonumber & & \left. + (d-1-2N)\ket{\nu_7} + 2\sqrt{d-N}\ket{\nu_8}\right), \\
\nonumber \hat U_m \ket{\nu_8} & = & \frac{1}{d+1}\left(2 \ket{\nu_2} + 2\sqrt{N-1}\ket{\nu_4} + \right. \\ 
\label{search:U:I} & & \left. + 2\sqrt{d-N}\ket{\nu_7} +  (1-d)\ket{\nu_8}\right) .
\end{eqnarray}
Hence, $\cal I$ is indeed an invariant subspace of the SA and the state of the walk $\ket{\phi(t)}$ remains in $\cal I$ for all $t$. Its evolution is determined by the eigenvalues and eigenvectors of $\hat U_m$, which is in the invariant subspace represented by an $8\times 8$ unitary matrix with matrix elements given by (\ref{search:U:I}). While it is possible to diagonalize this matrix analytically, the procedure is rather onerous and the resulting expressions are quite lengthy. Nevertheless, the analysis can be considerably simplified in the limit of a large graph, i.e., when $N\to\infty$ and $M\to\infty$. As we can see from the expansion (\ref{initInvariant}) for large $N$ and $M$ the initial state of the algorithm tends to $\ket{\nu_7}$. Hence, the only eigenvectors of the evolution operator $\hat U_m$, which remain relevant in the asymptotic limit, are those which have non-vanishing  {overlap with $\ket{\nu_7}$.} It turns out that there are only three such states and their asymptotic form is given by
\begin{eqnarray}
\nonumber \ket{\psi_1} & = & \frac{1}{\sqrt{2}} (\ket{\nu_7} - \ket{\nu_1}), \\
\label{search:evec} \ket{\psi_2^{(\pm)}} & = & \frac{1}{2}(\ket{\nu_1} + \ket{\nu_7} \pm i(\ket{\nu_5} - \ket{\nu_2})) .
\end{eqnarray}
 {It can be shown that for the other eigenvectors of $\hat U_m$ the overlap with the initial state decreases at least as $O(1/\sqrt{NM})$. Let us turn to the eigenvalues.} The eigenvector $\ket{\psi_1}$ corresponds to $\lambda_1 = 1$. For $\ket{\psi_2^{(\pm)}}$ the eigenvalues have the form 
$$
\lambda_2^{(\pm)} = e^{\pm i \omega_2}.
$$
From the characteristic polynomial of $\hat U_m$ we find that $\cos\omega_{2}$ is given by the largest root of the quadratic equation
\begin{eqnarray}
\nonumber 
x^2 - \left(1 - \frac{N}{d+1}\right)x - \frac{(d+1) (N-2)+N-3}{{(d+1)}^2} = 0 .
\end{eqnarray}
This leads us to
\begin{eqnarray}
\nonumber \omega_{2} & = & \arccos\left(1 - \frac{1+ NM - \sqrt{N^2 M^2 - 6 N M + 4N + 5 } }{2{(d+1)}} \right) \\
\label{search:omega} & \approx & \frac{2}{\sqrt{NM}}.
\end{eqnarray}
From the relations (\ref{search:evec}) we express the initial and the target state of the SA in terms of the eigenvectors of the evolution operator as
\begin{eqnarray}
\nonumber \ket{\Omega} & = & \frac{1}{\sqrt{2}} \ket{\psi_1} + \frac{1}{2}\left(\ket{\psi_2^{(+)}} + \ket{\psi_2^{(-)}}\right), \\
\ket{\nu_1}  & = & -\frac{1}{\sqrt{2}} \ket{\psi_1} + \frac{1}{2}\left(\ket{\psi_2^{(+)}} + \ket{\psi_2^{(-)}}\right) .
\label{search:target}
\end{eqnarray}
The state after $t$ iterations of the SA reads
\begin{equation}
    \ket{\phi(t)} = \frac{1}{\sqrt{2}} \ket{\psi_1} + \frac{1}{2}\left(e^{i\omega_2 t}\ket{\psi_2^{(+)}} + e^{-i\omega_2 t}\ket{\psi_2^{(-)}}\right).
\label{sa:state:t}
\end{equation}
The success probability (\ref{searchProbab}) of the SA after $t$ steps can be expressed in the form
\begin{equation}
\label{search:p}
P_m(t) = |\braket{\nu_1}{\phi(t)}|^2 + |\braket{\nu_2}{\phi(t)}|^2 .    
\end{equation}
From (\ref{sa:state:t}) and (\ref{search:target}) we see that the probability to find the walker in the target state $\ket{\nu_1}$ is given by
\begin{equation}
\label{search:loop}
    |\braket{\nu_1}{\phi(t)}|^2 = \sin^4  \left(\frac{\omega_2 t}{2}\right) .
\end{equation}
The probability to find the walker in the state $\ket{\nu_2}$ reads
\begin{equation}
|\braket{\nu_2}{\phi(t)}|^2 = \frac{1}{4} \sin^2\left(\omega_2 t\right) .
\label{search:ew}
\end{equation}
Put together we obtain the overall success probability of the SA
\begin{equation}
\label{search:prob:complete}
    P_m(t) = \sin^2  \left(\frac{\omega_2 t}{2}\right) .
\end{equation}
We see that for $t = \frac{\pi}{\omega_2}$ the state of the SA is very close to the target state $\ket{\nu_1}$. Hence, the number of steps needed to find the marked vertex with probability close to one is given by
\begin{equation}
\label{search:tmax}
    T = \frac{\pi}{\omega_2} \approx \frac{\pi \sqrt{NM}}{2} + O\left(\frac{1}{\sqrt{NM}}\right). 
\end{equation}

\begin{figure}
    \centering
     \includegraphics[width=0.45\textwidth]{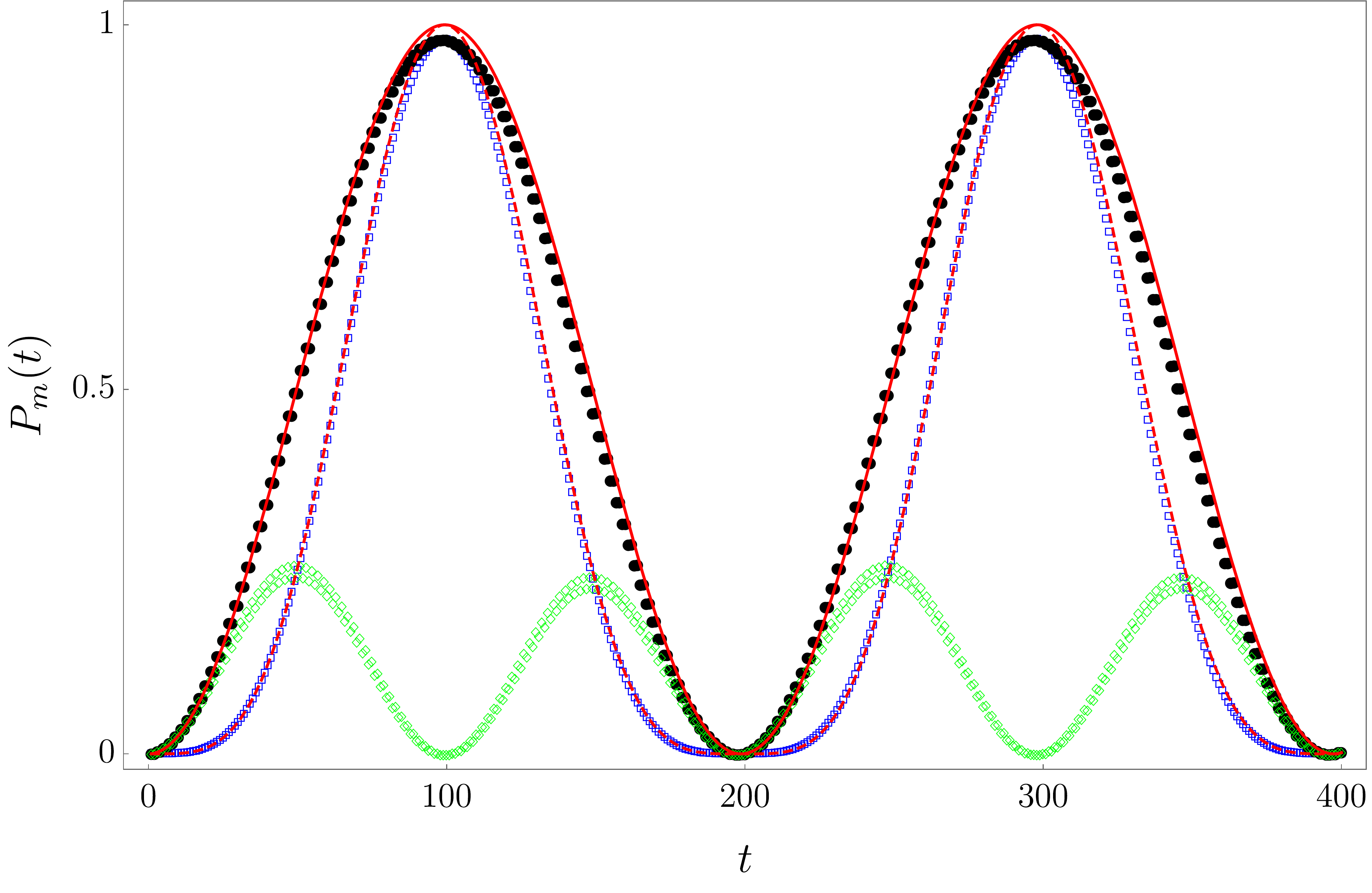}
    \caption{Overall success probability of SA (black dots) and the probability to find the walker in the target state $\ket{\nu_1}$ (blue squares) as a function of the number of steps $t$ for $N=40$ and $M=100$. Red curves correspond to the analytical results (dashed line to eq. (\ref{search:loop}) and full line to eq. (\ref{search:prob:complete}) ). Green diamonds denote the probability that the walker is on the marked vertex but not in the loop, which follows the curve (\ref{search:ew}). The success probability is close to one after $T\approx 100$ steps, in accordance with (\ref{search:tmax}).}
    \label{fig:search}
\end{figure}

For illustration we plot in Figure~\ref{fig:search} the probability to find the marked vertex (\ref{search:prob:complete}) as a function of the number of steps for a graph with $N=40$ and $M=100$.

\begin{figure}
    \centering
     \includegraphics[width=0.45\textwidth]{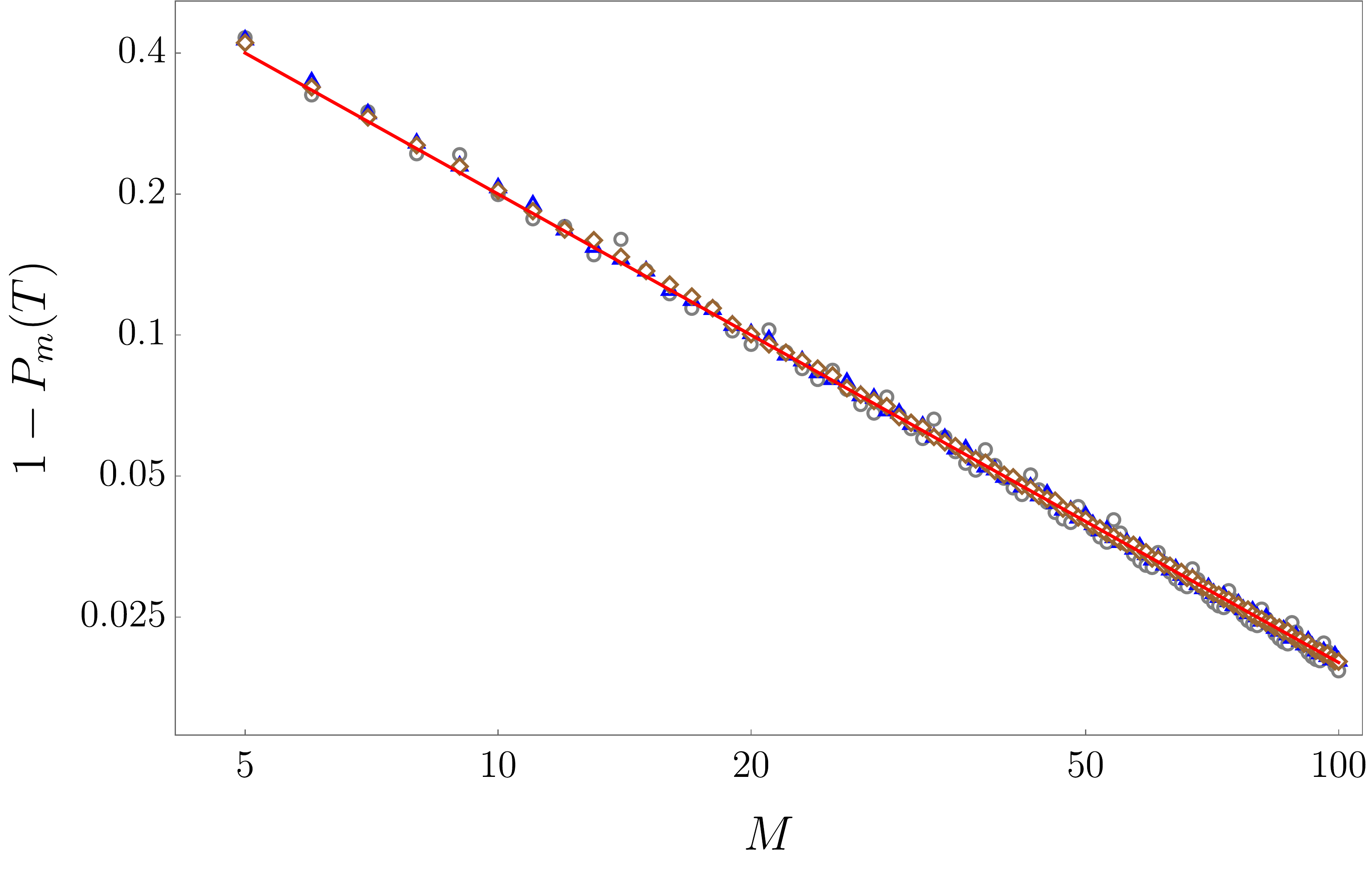}
    \caption{ {Overall success probability of SA as a function of the number of partitions $M$ for $N=10$ (gray circles), $N=50$ (blue triangles) and $N=100$ (brown diamonds). For a given $N$ and $M$ we evaluate numerically the evolution of SA for the optimal number of steps $T$ given by (\ref{search:tmax}) and determine $P_m(T)$ from the formula (\ref{search:p}). To unravel the scaling of the success probability we plot $1-P_m(T)$ on the log-log scale. Independent of the value of $N$, the sets of data-points fit well onto the $1/M$ slope indicated by the red line.}}
    \label{fig:search:nm}
\end{figure}

 {We note that the result (\ref{search:prob:complete}) holds in the limit of large $N$ and $M$. To investigate how quickly does the success probability at the optimal time (\ref{search:tmax}) approaches unity we performed numerical simulations for various values of $N$ and $M$. The simulations indicate that the success probability is essentially independent of $N$ and with $M$ it scales according to
$$
P_m(T) = 1 - O\left(\frac{1}{M}\right).
$$
The results are illustrated in Figure~\ref{fig:search:nm}.
}

\section{State transfer}
\label{sec:4}

\begin{figure}[ht]
 \centering
 \includegraphics[width=0.45\textwidth]{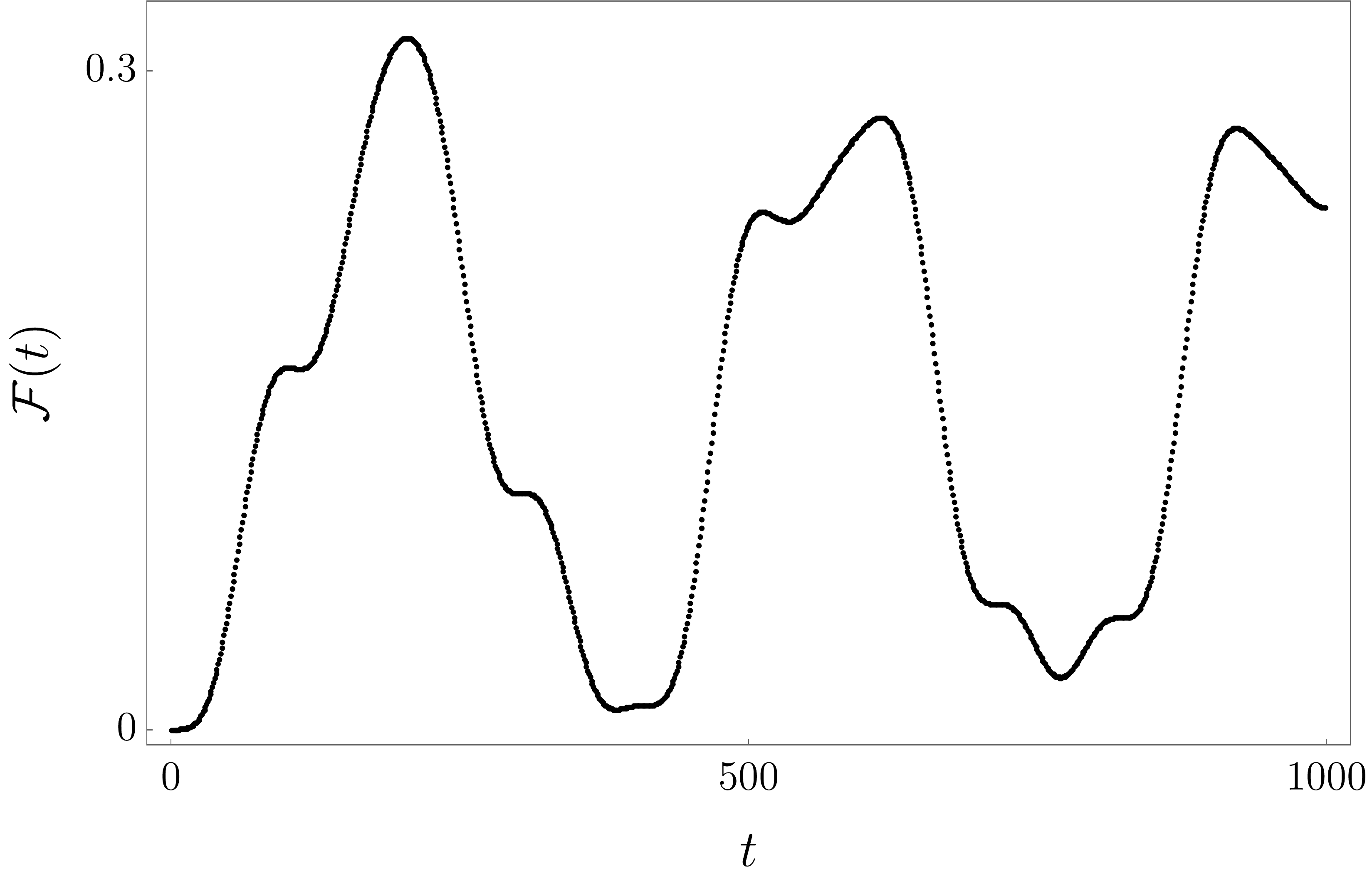}
 \caption{The evolution of the fidelity ${\cal F}$ of the state transfer during $1000$ steps for $N=40$ and $M=100$. Sender and receiver are in the same part. The initial state $\ket{s}$ is the equal weight superposition on the sender vertex $\ket{\Omega_s}$. We see that the fidelity does not grow over $0.35$.}
 \label{fig:same:ew}
\end{figure}

We now turn to the analysis of the STA. There are two possible configurations --- the sender and the receiver are in the same partition or not. Numerical simulations indicate that for the graph without loops the STA does not work well. When the sender and the receiver are in the same partition the fidelity does not surpass 0.25. For the second configuration the first maximum of fidelity tends to 0.8. Hence, we turn to the graph with loops. Choosing the initial state of the sender as the equal weight superposition (\ref{localSuperposition}), i.e., $\ket{s} = \ket{\Omega_s}$, the numerical simulation reveals that the fidelity tends to 1 for the second configuration. However, when the sender and the receiver are in the same partition the fidelity is still limited. As we show in Figure~\ref{fig:same:ew} it does not surpass 0.35.  {A careful analysis would reveal that the culprit are two orthogonal eigenvectors of the evolution operator of the STA corresponding to the eigenvalue -1, one having a large overlap with $\ket{\Omega_s}$ and the other one with $\ket{\Omega_r}$. In the second configuration this does not happen, as both $\ket{\Omega_s}$ and $\ket{\Omega_r}$ have overlaps with the same eigenvectors of the evolution operator of the STA. Hence, the absence of an edge between the sender and the receiver vertex in the first configuration significantly limits the achievable fidelity when we use the equal weight superposition state $\ket{\Omega_s}$ as the initial state of STA.}

We show that the fidelity of STA in the first configuration can be improved considerably by choosing the initial state $\ket{s}$ as the loop on the sender vertex. Moreover, we prove that this initial state works well also in the second configuration. In both configurations the walker will be with high probability transferred to the loop at the receiver vertex. We denote this receiver state as $\ket{r}$.

\subsection{Sender and receiver in the same partition}
\label{sec:4a}

Let us first consider the case when the sender and the receiver are in the same partition of the graph {, i.e., they are not connected directly by an edge}. Without loss of generality we consider that they are in the first one.

We begin by constructing the basis of the exact invariant subspace. The procedure is similar to the one for the SA, but we have to consider two marked vertices corresponding to the sender $s$ and the receiver $r$. Hence, the basis states of the form (\ref{search:nu1}), (\ref{search:nu2}), (\ref{search:nu5}) will appear twice - once for $m=s$ and once for $m=r$. In the end we find the following 11 basis vectors
\begin{eqnarray}
\nonumber \ket{\nu_1} & = & \ket{s_1,s_1} , \\
\nonumber \ket{\nu_2} & = & \frac{1}{\sqrt{d}}\sum_{\alpha=2}^M\sum_{j=1}^N\ket{s_1,j_\alpha} , \\
\nonumber \ket{\nu_3} & = & \ket{r_1,r_1} , \\
\nonumber \ket{\nu_4} & = & \frac{1}{\sqrt{d}}\sum_{\alpha=2}^M\sum_{j=1}^N\ket{r_1,j_\alpha} , \\
\nonumber \ket{\nu_5} & = & \frac{1}{\sqrt{N-2}}\sum_{j \neq s,r}^N\ket{j_1,j_1} , \\
\nonumber \ket{\nu_6} & = & \frac{1}{\sqrt{d(N-2)}}\sum_{j \neq s,r}^N\sum_{\alpha=2}^M\sum_{k=1}^N\ket{j_1,k_\alpha} , \\
\nonumber \ket{\nu_7} & = & 
\frac{1}{\sqrt{d}}\sum_{\alpha=2}^M\sum_{j=1}^N\ket{j_\alpha,s_1} , \\
\nonumber \ket{\nu_8} & = & 
\frac{1}{\sqrt{d}}\sum_{\alpha=2}^M\sum_{j=1}^N\ket{j_\alpha,r_1} , \\
\nonumber \ket{\nu_9} & = & 
\frac{1}{\sqrt{d(N-2)}}\sum_{j \neq s,r}^N\sum_{\alpha=2}^M\sum_{k=1}^N\ket{k_\alpha,j_1} , \\
\nonumber \ket{\nu_{10}} & = & \frac{1}{\sqrt{d(d-N)}}\sum_{\alpha=2}^M\sum_{\beta=2,\beta\neq\alpha}^M\sum_{j,k=1}^N\ket{j_\alpha,k_\beta} , \\
\label{same:basis} \ket{\nu_{11}} & = & \frac{1}{\sqrt{d}}\sum_{\alpha=2}^M\sum_{j}^N\ket{j_\alpha,j_\alpha}
\end{eqnarray}
Let us denote the subspace spanned by these vectors as $\cal I$. Clearly, it contains the initial and the desired target state of the STA ($\ket{s} = \ket{\nu_1}$ and $\ket{r} = \ket{\nu_3}$). It can be shown by direct calculation that it is closed under the action of $\hat U_{s,r}$. However, we will not provide the expression of $\hat U_{s,r}$ in this basis because $\cal I$ can be divided further into two invariant subspaces. This comes from a fact that the evolution of the STA is invariant with respect to the exchange of the sender and the receiver vertex. Let us denote the operator corresponding to this symmetry as $\hat P$. Clearly, it holds that 
$$
\hat P^2 = \hat I, \quad [\hat P, \hat U_{s,r}] = 0.
$$
$\hat P$ acts on the basis states (\ref{same:basis}) as
\begin{eqnarray}
\nonumber \hat P \ket{\nu_1} & = & \ket{\nu_3}, \quad \hat P \ket{\nu_2} = \ket{\nu_4}, \quad \hat P \ket{\nu_7} = \ket{\nu_8}, \\
\nonumber \hat P \ket{\nu_j} & = & \ket{\nu_j}, \quad j = 5,6,9,10,11 .
\end{eqnarray}
Since $\hat P$ commutes with $\hat U_{s,r}$, they have common eigenvectors. From $\hat{P}^2=\hat{I}$ we see that the spectrum of $\hat{P}$ consists of two eigenvalues $1$ and $-1$. Hence, the invariant subspace ${\cal I}$ can be split into two subspaces ${\cal I}_+$ and ${\cal I}_-$ which correspond to eigenvalues $1$ and $-1$ of the operator $\hat{P}$. Basis of the invariant subspace ${\cal I}_+$ is spanned by eigenstates denoted as $\ket{\sigma_i}$, $i=1,\ldots,8$ and it has the following form
\begin{eqnarray}
\nonumber \ket{\sigma_1} & = & \frac{1}{\sqrt{2}}\left(\ket{\nu_1}+\ket{\nu_3}\right) , \\
\nonumber \ket{\sigma_2} & = & \frac{1}{\sqrt{2}}\left(\ket{\nu_2}+\ket{\nu_4}\right) , \\
\nonumber \ket{\sigma_3} & = & \ket{\nu_5} , \\
\nonumber \ket{\sigma_4} & = & \ket{\nu_6} , \\
\nonumber \ket{\sigma_5} & = & \frac{1}{\sqrt{2}}\left(\ket{\nu_7}+\ket{\nu_8}\right) , \\
\nonumber \ket{\sigma_6} & = & \ket{\nu_9} , \\
\nonumber \ket{\sigma_7} & = & \ket{\nu_{10}} , \\
\nonumber \ket{\sigma_8} & = & \ket{\nu_{11}}
\end{eqnarray}
Note that if we perform SA for two marked vertices instead of STA, the subspace ${\cal I}_+$ would be invariant with respect to the search. This is due to the fact that the initial state of SA algorithm lies within this subspace. Basis of the invariant subspace ${\cal I}_-$ is spanned by eigenstates denoted as $\ket{\tau_i}$, $i=1,\ldots,3$ and it has the following form
\begin{eqnarray}
\nonumber \ket{\tau_1} & = & \frac{1}{\sqrt{2}}\left(\ket{\nu_1}-\ket{\nu_3}\right) , \\
\nonumber \ket{\tau_2} & = & \frac{1}{\sqrt{2}}\left(\ket{\nu_2}-\ket{\nu_4}\right) , \\
\nonumber \ket{\tau_3} & = & \frac{1}{\sqrt{2}}\left(\ket{\nu_7}-\ket{\nu_8}\right)
\end{eqnarray}
This subspace is needed only in STA since the initial state of SA is orthogonal to this subspace.

The sender and the receiver states in the new basis read
\begin{eqnarray}
\nonumber \ket{s} & = & \frac{1}{\sqrt{2}}\left(\ket{\sigma_1} + \ket{\tau_1} \right), \\
\label{transfer:same:sr} \ket{r} & = & \frac{1}{\sqrt{2}}\left(\ket{\sigma_1} - \ket{\tau_1} \right) .
\end{eqnarray}
The evolution operator in the new basis is block diagonal, i.e., ${\cal I}_i$ are the invariant subspaces of $\hat U_{s,r}$. We find the following relations for the basis states of ${\cal I}_+$
\begin{widetext}
\begin{eqnarray}
\nonumber \hat U_{s,r} \ket{\sigma_1} & = & \frac{1}{d+1}\left((d-1)\ket{\sigma_1} - 2\sqrt{d}\ket{\sigma_5}\right), \\
\nonumber \hat U_{s,r} \ket{\sigma_2} & = & -\frac{1}{d+1}\left( 2\sqrt{d}\ket{\sigma_1} + (d-1)\ket{\sigma_5}\right), \\ 
\nonumber \hat U_{s,r} \ket{\sigma_3} & = & \frac{1}{d+1}\left((1-d)\ket{\sigma_3} + 2\sqrt{d}\ket{\sigma_6}\right), \\
\nonumber \hat U_{s,r} \ket{\sigma_4} & = & \frac{1}{d+1}\left(2\sqrt{d}\ket{\sigma_3} + (d-1)\ket{\sigma_6}\right), \\
\nonumber \hat U_{s,r} \ket{\sigma_5} & = & \frac{1}{d+1}\left( (3-d)\ket{\sigma_2} + 2\sqrt{2(N-2)}\ket{\sigma_4} + 2\sqrt{2(d-N)}\ket{\sigma_7} + 2\sqrt{2}\ket{\sigma_8} \right), \\
\nonumber \hat U_{s,r} \ket{\sigma_6} & = & \frac{1}{d+1}\left( 2\sqrt{2(N-2)}\ket{\sigma_2} -\left(d-2N+5\right) \ket{\sigma_4} + 2\sqrt{(d-N)(N-2)}\ket{\sigma_7} + 2\sqrt{N-2}\ket{\sigma_8} \right), \\
\nonumber \hat U_{s,r} \ket{\sigma_7} & = & \frac{1}{d+1}\left( 2\sqrt{2(d-N)}\ket{\sigma_2} + 2\sqrt{(d-N)(N-2)} \ket{\sigma_4} + \left(d-2N-1\right)\ket{\sigma_7} + 2\sqrt{d-N}\ket{\sigma_8} \right), \\
\hat U_{s,r} \ket{\sigma_8} & = & \frac{1}{d+1} \left( 2\sqrt{2}\ket{\sigma_2} + 2\sqrt{N-2}\ket{\sigma_4} + 2\sqrt{d-N}\ket{\sigma_7} - \left(d-1\right)\ket{\sigma_8} \right).
\end{eqnarray}
\end{widetext}
In the second subspace ${\cal I}_-$ the evolution operator acts according to
\begin{eqnarray}
\nonumber  \hat U_{s,r} \ket{\tau_1} & = & \frac{1}{d+1} \left( \left(d-1\right)\ket{\tau_1} - 2\sqrt{d}\ket{\tau_3} \right), \\
\nonumber \hat U_{s,r} \ket{\tau_2} & = & -\frac{1}{d+1} \left( 2\sqrt{d}\ket{\tau_1} + \left(d-1\right) \ket{\tau_3} \right), \\
\hat U_{s,r} \ket{\tau_3} & = & -\ket{\tau_2} . 
\end{eqnarray}
Let us now investigate the dynamics of the STA in the limit of a large graph. We denote by $\hat U_\pm$ the restriction of $\hat U_{s,r}$ on ${\cal I}_\pm$, and determine the spectrum and eigenvectors of these operators. For $\hat U_+$ the results are similar to those for the SA. In the subspace ${\cal I}_+$ three relevant eigenvectors remain,  {as for the others the overlap with $\ket{\sigma_1}$ tends to zero at least as $O(1/\sqrt{NM})$. The limit form of the relevant eigenvectors is given by}
\begin{eqnarray}
\nonumber \ket{\psi_1} & = & \sqrt{\frac{2}{3}}\ket{\sigma_1} - \frac{1}{\sqrt{3}}\ket{\sigma_7}, \\
\label{same:evec1} \ket{\psi_2^{(\pm)}} & = &  \frac{1}{\sqrt{6}} \ket{\sigma_1} + \frac{1}{\sqrt{3}}\ket{\sigma_7}  \pm\frac{i}{2}(\ket{\sigma_5} - \ket{\sigma_2}) .
\end{eqnarray}
The eigenvector $\ket{\psi_1}$ corresponds to $\lambda_1 = 1$. In the case of $\ket{\psi_2^{(\pm)}}$ the eigenvalues have the form
$$
\lambda_2^{(\pm)} = e^{\pm i\omega_2}.
$$
From the characteristic polynomial of $\hat U_+$ we find that  $\cos\omega_2$ is the largest root of the quadratic equation
$$
x^2 - \left(1 - \frac{N}{d+1}\right)x - \frac{{(d+1)}(N-3) + N - 5}{{(d+1)}^2}  = 0,
$$
which leads us to
\begin{eqnarray}
\nonumber \omega_{2} & = & \arccos\left(1 - \frac{1+ N M - \sqrt{N^2 M^2 - 10 NM + 8 N + 9}}{2{(d+1)}}\right) \\
\label{same:omega2} & \approx & \sqrt{\frac{6}{NM}} .
\end{eqnarray}
In the subspace $\cal I_-$ there are two additional relevant eigenvectors in the asymptotic limit
\begin{equation}
\label{same:evec2}
\ket{\psi_3^{(\pm)}} = \frac{1}{\sqrt{2}}\ket{\tau_1} \pm \frac{i}{2}(\ket{\tau_2} - \ket{\tau_3}) .     
\end{equation}
 {For the last eigenvector of $\hat U_-$ the overlap with the state $\ket{\tau_1}$ behaves like $O(1/\sqrt{NM})$. The relevant eigenvalues} have the form
\begin{eqnarray}
\nonumber \lambda_3^{(\pm)} & = & e^{\pm i\omega_3},
\end{eqnarray}
with
\begin{equation}
\label{same:omega3}
    \omega_3 = \arccos\left(1-\frac{1}{{d+1}}\right) \approx \sqrt{\frac{2}{NM}} .
\end{equation}
Using the results (\ref{same:evec1}), (\ref{same:evec2}) we see that for a large graph the sender and the receiver states (\ref{transfer:same:sr}) can be decomposed into the eigenvectors of the evolution operator $\hat U_{s,r}$ according to
\begin{eqnarray}
\nonumber \ket{s} & = & \frac{1}{\sqrt{3}} \ket{\psi_1} + \frac{1}{\sqrt{12}} \left(\ket{\psi_2^{(+)}} + \ket{\psi_2^{(-)}}\right) + \\
\nonumber & & + \frac{1}{2} \left(\ket{\psi_3^{(+)}} + \ket{\psi_3^{(-)}}\right) , \\
\nonumber \ket{r} & = & \frac{1}{\sqrt{3}} \ket{\psi_1} + \frac{1}{\sqrt{12}} \left(\ket{\psi_2^{(+)}} + \ket{\psi_2^{(-)}}\right) - \\
\nonumber & & - \frac{1}{2} \left(\ket{\psi_3^{(+)}} + \ket{\psi_3^{(-)}}\right) .
\end{eqnarray}
Hence, the evolution of STA takes place in a five dimensional subspace
\begin{eqnarray}
\nonumber   \ket{\phi(t)} & = & \frac{1}{\sqrt{3}} \ket{\psi_1} + \\
\nonumber & &  + \frac{1}{\sqrt{12}} \left(e^{i \omega_2 t}\ket{\psi_2^{(+)}} + e^{-i \omega_2 t}\ket{\psi_2^{(-)}}\right) + \\
\label{same:evol} & & + \frac{1}{2} \left(e^{i \omega_3 t}\ket{\psi_3^{(+)}} + e^{-i \omega_3 t}\ket{\psi_3^{(-)}}\right) .
\end{eqnarray}
The fidelity of STA can be written as a sum 
\begin{equation}
\label{same:f}
{\cal F}(t) = |\braket{r}{\phi(t)}|^2 + |\braket{\nu_4}{\phi(t)}|^2,
\end{equation}
of probabilities that the walker is in the receiver state $\ket{r} = \ket{\nu_3}$ corresponding to the loop, or at the receiver vertex but not in the loop, i.e., the state $\ket{\nu_4}$. From the relations (\ref{same:evec1}), (\ref{same:evec2}) and (\ref{same:evol}) we find that these probabilities are given by
\begin{eqnarray}
\label{same:fid:r} |\braket{r}{\phi(t)}|^2 & = & \frac{1}{36}\left(2 + \cos\left(\omega_2 t\right)  - 3\cos\left(\omega_3 t\right)\right)^2 , \\
\label{same:fid:nr} |\braket{\nu_4}{\phi(t)}|^2 & = & \frac{1}{24}\left(\sin(\omega_2 t) - \sqrt{3} \sin(\omega_3 t)\right)^2 .
\end{eqnarray}
From the relations (\ref{same:omega2}), (\ref{same:omega3}) we see that the frequencies are not harmonic, since for a large graph 
$$
\omega_2 = \sqrt{3}\omega_3 .
$$
The overall fidelity of STA is then given by
\begin{eqnarray}
\nonumber {\cal F}(t) & = & \frac{1}{36}\left(2 + \cos\left(\sqrt{3}\omega_3 t\right)  - 3\cos\left(\omega_3 t\right)\right)^2 + \\
\label{same:fid} & & + \frac{1}{24}\left(\sin(\sqrt{3}\omega_3 t) - \sqrt{3} \sin(\omega_3 t)\right)^2 .
\end{eqnarray}
The fidelity of STA will not reach one exactly. However, for a large graph the first maximum of fidelity reaches the value
\begin{equation}
\label{same:f:1}
    {\cal F}_{1} \approx 0.94 .
\end{equation}
The number of steps required to reach the first maximum is approximately given by
\begin{equation}
    T^{(st)}\approx 2.39 \sqrt{N M} .
\label{same:tmax}
\end{equation}
At this time the walker is with high probability in the receiver state $\ket{r}$.

\begin{figure}
    \centering
    \includegraphics[width=0.45\textwidth]{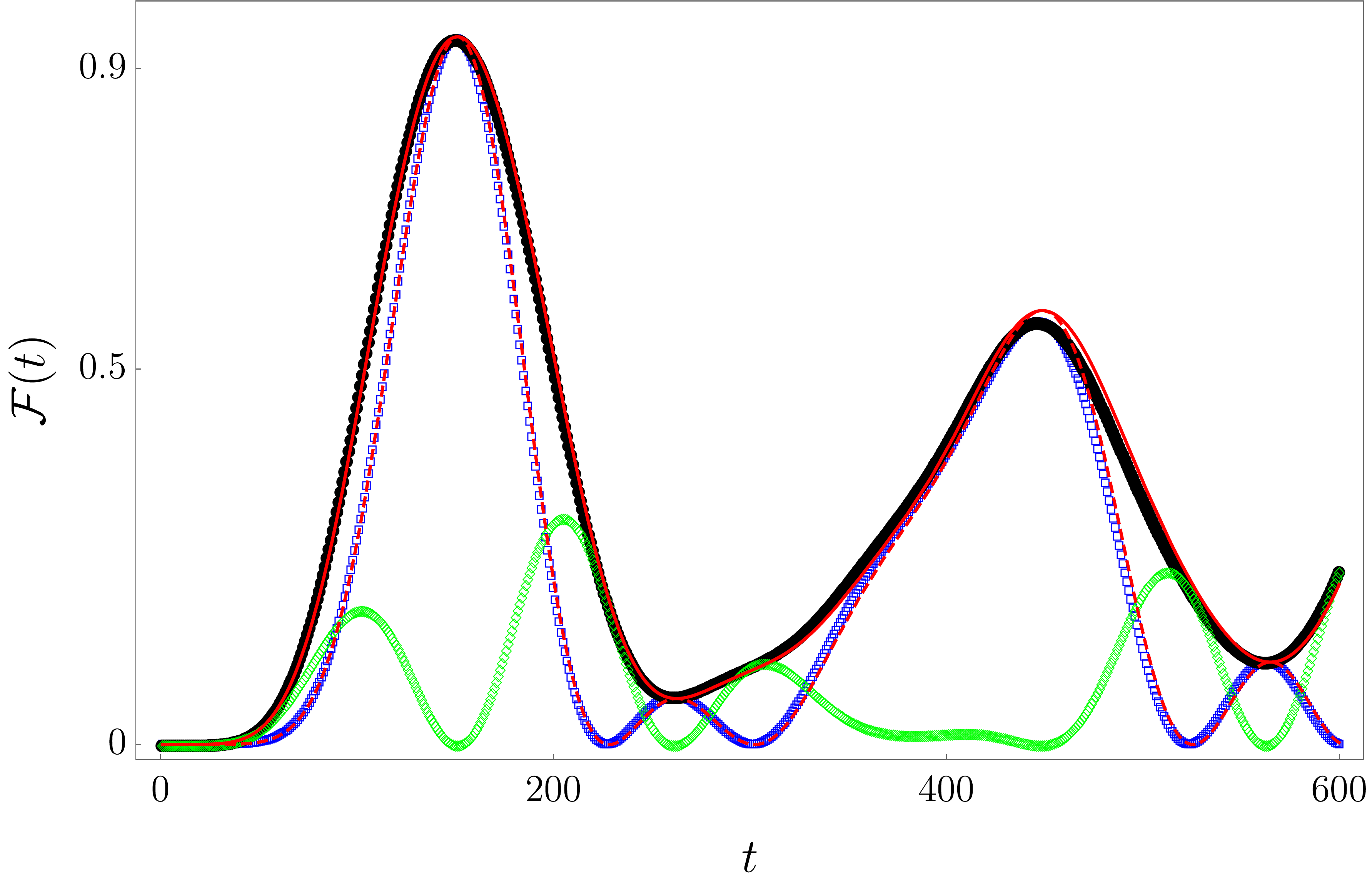}
    \caption{Fidelity of the state transfer as a function of the number of steps $t$ for $N=40$ and $M=100$. The sender and the receiver vertices are in the same partition.  {Black dots are obtained from the numerical simulation, the full red curve corresponds to (\ref{same:fid}).} Since the frequencies (\ref{same:omega2}), (\ref{same:omega3}) are not integer multiples the fidelity behaves an-harmonically. At the time of the first maximum (\ref{same:tmax}) the walker is with high probability in the receiver state $\ket{r}$, which is depicted by the blue squares. The probability to be in the receiver state follows the curve (\ref{same:fid:r})  {represented by the red dashed curve}. The green diamonds correspond to the probability that the walker is at the marked vertex but not in the loop, which follows the curve (\ref{same:fid:nr}). }
    \label{fig:same}
\end{figure}

For illustration we show in Figure~\ref{fig:same} the evolution of fidelity for a graph with $N=40$ and $M=100$.

\begin{figure}
    \centering
     \includegraphics[width=0.45\textwidth]{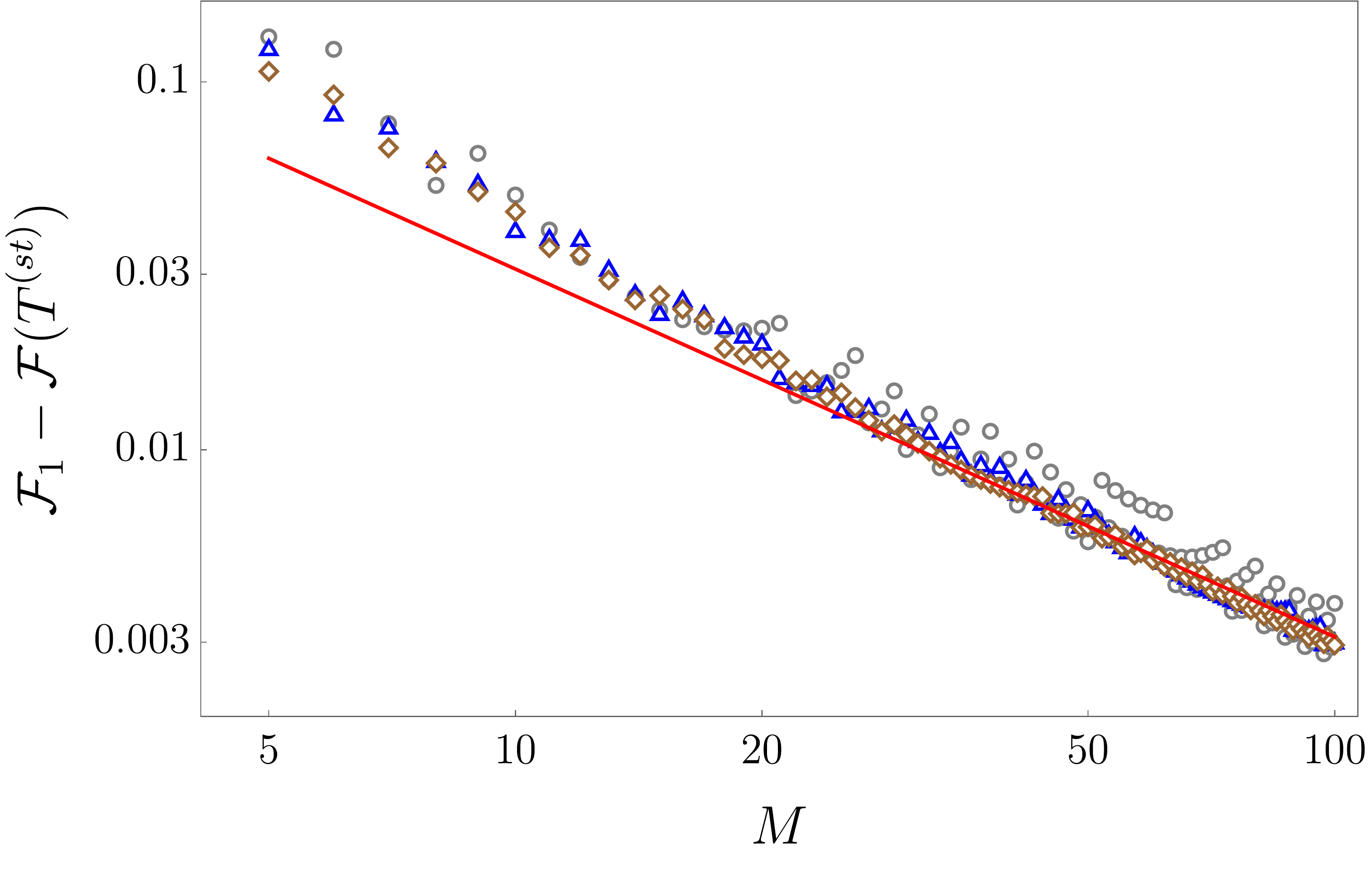}
    \caption{ {Overall fidelity of STA as a function of the number of partitions $M$ for $N=10$ (gray circles), $N=50$ (blue triangles) and $N=100$ (brown diamonds). The sender and the receiver vertices are in the same partition. For a given $N$ and $M$ we evaluate numerically the evolution of STA for the number of steps $T^{(st)}$ needed to reach the first maximum (\ref{same:tmax}) and determine ${\cal F}(T^{(st)})$ according to (\ref{same:f}). To unravel the scaling of the fidelity we plot ${\cal F}_1-{\cal F}(T^{(st)})$ on the log-log scale. The data-points follow the $1/M$ slope indicated by the red line, with almost no dependence on $N$.}}
    \label{fig:same:nm}
\end{figure}

 {The fidelity (\ref{same:f:1}) in the first maximum is reached in the limit of large $N$ and $M$. We have performed numerical simulations to investigate how quickly does the fidelity at the optimal time (\ref{same:tmax}) approaches the asymptotic value (\ref{same:f:1}). Similarly to the results for the SA, the simulations indicate that the fidelity is essentially independent of $N$ and with $M$ it scales according to
$$
{\cal F}(T^{(st)}) = {\cal F}_1 - O\left(\frac{1}{M}\right).
$$
The results are illustrated in Figure~\ref{fig:same:nm}.
}

\subsection{Sender and receiver in different partitions}
\label{sec:4b}

Let us now turn to the case when the sender and the receiver are in different parts of the complete M-partite graph. Without loss of generality we label the partition containing the sender as 1 and the partition with the receiver as 2.

The construction of the basis of the invariant subspace $\cal I$ is more involved, since we have to consider the second partition with the receiver vertex separately from the rest of the graph. We begin with the states starting at the sender vertex
\begin{eqnarray}
\nonumber \ket{\nu_1} & = & \ket{s_1,s_1} , \\
\nonumber \ket{\nu_2} & = & \ket{s_1,r_2} , \\
\nonumber \ket{\nu_3} & = & \frac{1}{\sqrt{N-1}}\sum_{j\neq r}^N\ket{s_1,j_2} , \\
\label{diff:basis:1} \ket{\nu_4} & = & \frac{1}{\sqrt{d-N}}\sum_{\alpha=3}^M\sum_{j=1}^N\ket{s_1,j_\alpha},
\end{eqnarray}
which correspond to the loop, edge from the sender to the receiver, equal weight superposition of all edges from the sender to the remaining vertices in the second partition, and edges to all remaining vertices. We repeat the same for the receiver vertex 
\begin{eqnarray}
\nonumber \ket{\nu_5} & = & \ket{r_2,r_2} , \\
\nonumber \ket{\nu_6} & = & 
\ket{r_2,s_1} , \\
\nonumber \ket{\nu_7} & = & \frac{1}{\sqrt{N-1}}\sum_{j\neq s}^N\ket{r_2,j_1} , \\
\label{diff:basis:2} \ket{\nu_8} & = & \frac{1}{\sqrt{d-N}}\sum_{\alpha=3}^M\sum_{j=1}^N\ket{r_2,j_\alpha} 
\end{eqnarray}
Next, we consider the same edges but starting at the non-marked vertices in the first partition and prepare the following superpositions
\begin{eqnarray}
\nonumber \ket{\nu_9} & = & \frac{1}{\sqrt{N-1}}\sum_{j\neq s}^N\ket{j_1,j_1} , \\
\nonumber \ket{\nu_{10}} & = & 
\frac{1}{\sqrt{N-1}}\sum_{j\neq s}^N\ket{j_1,r_2} , \\
\nonumber \ket{\nu_{11}} & = & \frac{1}{N-1}\sum_{j\neq s}^N\sum_{k\neq r}^N\ket{j_1,k_2} , \\
\label{diff:basis:3} \ket{\nu_{12}} & = & \frac{1}{\sqrt{(d-N)(N-1)}}\sum_{j\neq s}^N\sum_{\alpha=3}^M\sum_{k=1}^N\ket{j_1,k_\alpha} .
\end{eqnarray}
The same procedure is repeated in the second partition
\begin{eqnarray}
\nonumber \ket{\nu_{13}} & = & \frac{1}{\sqrt{N-1}}\sum_{j\neq r}^N\ket{j_2,j_2} , \\
\nonumber \ket{\nu_{14}} & = & 
\frac{1}{\sqrt{N-1}}\sum_{j\neq r}^N\ket{j_2,s_1} , \\
\nonumber \ket{\nu_{15}} & = & 
\frac{1}{N-1}\sum_{j\neq r}^N\sum_{k\neq s}^N\ket{j_2,k_1} , \\
\label{diff:basis:4} \ket{\nu_{16}} & = & \frac{1}{\sqrt{(d-N)(N-1)}}\sum_{j\neq r}^N\sum_{\alpha=3}^M\sum_{k=1}^N\ket{j_2,k_\alpha} .
\end{eqnarray}
Next, we consider the edges leading to the sender or the receiver vertex from the outside of the first two partitions
\begin{eqnarray}
\nonumber \ket{\nu_{17}} & = & 
\frac{1}{\sqrt{d-N}}\sum_{\alpha=3}^M\sum_{j=1}^N\ket{j_\alpha,s_1} , \\ 
\label{diff:basis:5} \ket{\nu_{18}} & = & 
\frac{1}{\sqrt{d-N}}\sum_{\alpha=3}^M\sum_{j=1}^N\ket{j_\alpha,r_2}.
\end{eqnarray}
Then we add states corresponding to edges leading to the non-marked vertices in the first or the second partition from the outside
\begin{eqnarray}
\nonumber \ket{\nu_{19}} & = & 
\frac{1}{\sqrt{(d-N)(N-1)}}\sum_{j\neq s}^N\sum_{\alpha=3}^M\sum_{k=1}^N\ket{k_\alpha,j_1}, \\
\label{diff:basis:6} \ket{\nu_{20}} & = & 
\frac{1}{\sqrt{(d-N)(N-1)}}\sum_{j\neq r}^N\sum_{\alpha=3}^M\sum_{k=1}^N\ket{k_\alpha,j_2}.
\end{eqnarray}
Finally, we consider all edges between vertices in the rest of the graph, and all remaining loops 
\begin{eqnarray}
\nonumber \ket{\nu_{21}} & = & \frac{1}{\sqrt{N(d-N)(M-3)}}\sum_{\alpha=3}^M\sum_{\beta=3,\beta\neq\alpha}^M\sum_{j,k=1}^N\ket{j_\alpha,k_\beta} , \\
\label{diff:basis:7} \ket{\nu_{22}} & = & \frac{1}{\sqrt{d-N}}\sum_{\alpha=3}^M\sum_{j=1}^N\ket{j_\alpha,j_\alpha}
\end{eqnarray}
It can be shown by direct calculation that the 22 vectors (\ref{diff:basis:1})-(\ref{diff:basis:7}) constitute an invariant subspace of the STA. To proceed further we employ the symmetry $\hat P$ which switches the sender and the receiver partitions. Its action on the basis states $\ket{\nu_j}$ is given by
\begin{eqnarray}
\nonumber \hat P\ket{\nu_j} & = & \ket{\nu_{j+4}}, \quad j = 1,2,3,4,9,10,11,12, \\
\nonumber \hat P\ket{\nu_i} & = & \ket{\nu_{i+1}}, \quad i = 17, 19 ,\\
\nonumber \hat P\ket{\nu_k} & = & \ket{\nu_{k}}, \quad k = 21, 22 . 
\end{eqnarray}
Since $\hat P$ commutes with the evolution operator of the STA we can split $\cal I$ into subspaces ${\cal I}_\pm$ corresponding to eigenvalues $\pm 1$. Subspace ${\cal I}_+$ has dimension $12$ and it is spanned by the following eigenvectors of $\hat{P}$
\begin{eqnarray}
\nonumber \ket{\sigma_i} & = & \frac{1}{\sqrt{2}}\left(\ket{\nu_i}+\ket{\nu_{i+4}}\right) ,\quad i = 1,2,3,4 \\
\nonumber \ket{\sigma_j} & = & \frac{1}{\sqrt{2}}\left(\ket{\nu_{j+4}}+\ket{\nu_{j+8}}\right) ,\quad j = 5,6,7,8 \\
\nonumber \ket{\sigma_9} & = & \frac{1}{\sqrt{2}}\left(\ket{\nu_{17}}+\ket{\nu_{18}}\right) , \\
\nonumber \ket{\sigma_{10}} & = & \frac{1}{\sqrt{2}}\left(\ket{\nu_{19}}+\ket{\nu_{20}}\right) , \\
\nonumber \ket{\sigma_{11}} & = & \ket{\nu_{21}} , \\
\nonumber \ket{\sigma_{12}} & = & \ket{\nu_{22}}
\end{eqnarray}
Subspace ${\cal I}_-$ has dimension $10$ and it is spanned by the following eigenvectors of $\hat{P}$ corresponding to the eigenvalue $-1$
\begin{eqnarray}
\nonumber \ket{\tau_i} & = & \frac{1}{\sqrt{2}}\left(\ket{\nu_i} - \ket{\nu_{i+4}}\right) ,\quad i = 1,2,3,4 \\
\nonumber \ket{\tau_j} & = & \frac{1}{\sqrt{2}}\left(\ket{\nu_{j+4}}-\ket{\nu_{j+8}}\right) ,\quad j = 5,6,7,8 \\
\nonumber \ket{\tau_9} & = & \frac{1}{\sqrt{2}}\left(\ket{\nu_{17}}-\ket{\nu_{18}}\right) , \\
\nonumber \ket{\tau_{10}} & = & \frac{1}{\sqrt{2}}\left(\ket{\nu_{19}}-\ket{\nu_{20}}\right)
\end{eqnarray}
In the new basis the sender and the receiver states have the following form 
\begin{eqnarray}
\nonumber \ket{s} & = & \frac{1}{\sqrt{2}}\left(\ket{\sigma_1} + \ket{\tau_1}\right), \\
\ket{r} & = & \frac{1}{\sqrt{2}}\left(\ket{\sigma_1} - \ket{\tau_1}\right) .
\label{different:s:r}
\end{eqnarray}
\begin{widetext}
The evolution operator $\hat U_{s,r}$ is block diagonal. We find the following relations for the basis vectors of ${\cal I}_+$
\begin{eqnarray}
\nonumber \hat U_{s,r} \ket{\sigma_1} & = & \frac{1}{d+1} \left(\left(d-1\right)\ket{\sigma_1} - 2\ket{\sigma_2} - 2\sqrt{N-1}\ket{\sigma_6} - 2\sqrt{d-N}\ket{\sigma_9} \right), \\
\nonumber \hat U_{s,r} \ket{\sigma_2} & = & \frac{1}{d+1} \left( -2\ket{\sigma_1} + \left(d-1\right)\ket{\sigma_2} - 2\sqrt{N-1}\ket{\sigma_6} - 2\sqrt{d-N}\ket{\sigma_9} \right), \\
\nonumber \hat U_{s,r} \ket{\sigma_3} & = & \frac{1}{d+1} \left( - 2\sqrt{N-1}\ket{\sigma_1} - 2\sqrt{N-1}\ket{\sigma_2} + \left(d-2N+3\right)\ket{\sigma_6} - 2\sqrt{(d-N)(N-1)}\ket{\sigma_9} \right), \\
\nonumber \hat U_{s,r}\ket{\sigma_4} & = & \frac{1}{d+1} \left(- 2\sqrt{d-N}\ket{\sigma_1} - 2\sqrt{d-N}\ket{\sigma_2} -  2\sqrt{(d-N)(N-1)}\ket{\sigma_6} -\left(d-2N-1\right) \ket{\sigma_9} \right), \\
\nonumber \hat U_{s,r}\ket{\sigma_5} & = & \frac{1}{d+1} \left( 2\ket{\sigma_3} - \left(d-1\right)\ket{\sigma_5} + 2\sqrt{N-1}\ket{\sigma_7} + 2\sqrt{d-N}\ket{\sigma_{10}} \right),\\
\nonumber \hat U_{s,r}\ket{\sigma_6} & = & \frac{1}{d+1} \left( -\left(d - 1\right)\ket{\sigma_3} + 2\ket{\sigma_5} +  2\sqrt{N-1}\ket{\sigma_7} + 2\sqrt{d-N}\ket{\sigma_{10}} \right),\\
\nonumber \hat U_{s,r}\ket{\sigma_7} & = & \frac{1}{d+1} \left( 2\sqrt{N-1}\ket{\sigma_3} + 2\sqrt{N-1}\ket{\sigma_5} - \left(d-2N+3\right) \ket{\sigma_7} + 2\sqrt{(d-N)(N-1)}\ket{\sigma_{10}} \right),\\
\nonumber \hat U_{s,r}\ket{\sigma_8} & = & \frac{1}{d+1} \left( 2\sqrt{d-N}\ket{\sigma_3} + 2\sqrt{d-N}\ket{\sigma_5} + 2\sqrt{(d-N)(N-1)} \ket{\sigma_7} + \left(d-2N-1\right)\ket{\sigma_{10}} \right),\\
\nonumber \hat U_{s,r}\ket{\sigma_9} & = & \frac{1}{d+1} \left( -\left(d-3\right)\ket{\sigma_4} + 4\sqrt{N-1}\ket{\sigma_8} + 2\sqrt{2N(M-3)}\ket{\sigma_{11}} + 2\sqrt{2}\ket{\sigma_{12}} \right), \\
\nonumber \hat U_{s,r}\ket{\sigma_{10}} & = & \frac{1}{d+1} \left( 4\sqrt{N-1}\ket{\sigma_4} - \left(d-4N+5\right)\ket{\sigma_8} + 2\sqrt{2N(N-1)(M-3)}\ket{\sigma_{11}} + 2\sqrt{2(N-1)}\ket{\sigma_{12}} \right), \\
\nonumber \hat U_{s,r}\ket{\sigma_{11}} & = & \frac{1}{d+1} \left( 2\sqrt{2N(M-3)}\ket{\sigma_4} + 2\sqrt{2N(N-1)(M-3)}\ket{\sigma_8} + \left(d-4N-1\right)\ket{\sigma_{11}} + 2\sqrt{N(M-3)}\ket{\sigma_{12}} \right), \\
\nonumber \hat U_{s,r}\ket{\sigma_{12}} & = & \frac{1}{d+1} \left( 2\sqrt{2}\ket{\sigma_4} + 2\sqrt{2(N-1)}\ket{\sigma_8} + 2\sqrt{N(M-3)}\ket{\sigma_{11}} - \left(d-1\right)\ket{\sigma_{12}} \right).
\end{eqnarray}
The action of the evolution operator on the basis vectors of ${\cal I}_-$ reads
\begin{eqnarray}
\nonumber \hat U_{s,r}\ket{\tau_1} & = & \frac{1}{d+1} \left(\left(d-1\right)\ket{\tau_1} + 2\ket{\tau_2}  + 2\sqrt{N-1}\ket{\tau_6} - 2\sqrt{d-N}\ket{\tau_9} \right), \\
\nonumber \hat U_{s,r}\ket{\tau_2} & = &  \frac{1}{d+1} \left( - 2\ket{\tau_1} - \left(d-1\right)\ket{\tau_2}  + 2\sqrt{N-1}\ket{\tau_6} - 2\sqrt{d-N}\ket{\tau_9} \right), \\
\nonumber \hat U_{s,r}\ket{\tau_3} & = &  \frac{1}{d+1} \left( - 2\sqrt{N-1}\ket{\tau_1} + 2\sqrt{N-1}\ket{\tau_2}  - \left(d-2N+3\right)\ket{\tau_6} - 2\sqrt{(d-N)(N-1)}\ket{\tau_9} \right), \\
\nonumber \hat U_{s,r}\ket{\tau_4} & = &  \frac{1}{d+1} \left(- 2\sqrt{d-N}\ket{\tau_1} + 2\sqrt{d-N}\ket{\tau_2}  + 2\sqrt{(d-N)(N-1)}\ket{\tau_6} - \left(d-2N-1\right)\ket{\tau_9} \right), \\
\nonumber \hat U_{s,r}\ket{\tau_5} & = & \frac{1}{d+1} \left( -2\ket{\tau_3} - \left(d-1\right)\ket{\tau_5} - 2\sqrt{N-1}\ket{\tau_7} + 2\sqrt{d-N}\ket{\tau_{10}} \right), \\
\nonumber \hat U_{s,r}\ket{\tau_6} & = & \frac{1}{d+1} \left( \left(d-1\right)\ket{\tau_3}  + 2\ket{\tau_5} - 2\sqrt{N-1}\ket{\tau_7} + 2\sqrt{d-N}\ket{\tau_{10}} \right), \\
\nonumber \hat U_{s,r}\ket{\tau_7} & = & \frac{1}{d+1} \left( - 2\sqrt{N-1}\ket{\tau_3}  + 2\sqrt{N-1}\ket{\tau_5}  + \left(d-2N+3\right)\ket{\tau_7} + 2\sqrt{(d-N)(N-1)}\ket{\tau_{10}} \right), \\
\nonumber \hat U_{s,r}\ket{\tau_8} & = & \frac{1}{d+1} \left( - 2\sqrt{d-N}\ket{\tau_3}  + 2\sqrt{d-N}\ket{\tau_5} - 2\sqrt{(d-N)(N-1)}\ket{\tau_7} + \left(d-2N-1\right)\ket{\tau_{10}} \right), \\
\nonumber \hat U_{s,r}\ket{\tau_9} & = & - \ket{\tau_4} , \\
 \hat U_{s,r}\ket{\tau_{10}} & = & - \ket{\tau_8} .
\end{eqnarray}
\end{widetext}
To investigate the dynamics of STA in more detail we again turn to the limit of a large graph. We denote by $\hat U_\pm$ the restriction of $\hat U_{s,r}$ on ${\cal I}_\pm$. For $\hat U_+$ there are three eigenstates which have non-vanishing overlap with the  {state $\ket{\sigma_1}$}, namely
\begin{eqnarray}
\label{diff:evec1}  \ket{\psi_1} & = & \frac{\sqrt{3}}{2}\ket{\sigma_1} - \frac{1}{2\sqrt{3}}\ket{\sigma_2} -\frac{1}{\sqrt{6}} \ket{\sigma_{11}}, \\
\nonumber \ket{\psi_2^{(\pm)}} & = & \frac{1}{\sqrt{8}}(\ket{\sigma_1} + \ket{\sigma_2}) \pm \frac{i}{2}(\ket{\sigma_9} - \ket{\sigma_4}) + \frac{1}{2}\ket{\sigma_{11}}. 
\end{eqnarray}
 {Note that for the other eigenstates the overlap with $\ket{\sigma_1}$ decreases at least as $O(1/\sqrt{NM})$. Turning to the eigenvalues, we find that the} eigenvector $\ket{\psi_1}$ has eigenvalue $\lambda_1 = 1$. From the characteristic polynomial of $\hat U_+$ we find that the eigenvalues of $\ket{\psi_2^{(\pm)}}$ have the form $\lambda_2^{(\pm)} = e^{\pm i\omega_2}$, where $\cos\omega_2$ is the largest root of the cubic equation
\begin{eqnarray}
\nonumber 0 & = &  x^3 - \left(1-\frac{N+2}{d+1}\right)x^2 - \frac{{(d+1)}(N-1) - N + 5}{{(d+1)}^2}x + \\
& & + \frac{N M - 4}{{(d+1)}^2}.
\nonumber 
\end{eqnarray}
We find that it has the following asymptotic form
\begin{equation}
\label{different:omega2}
  \omega_2 \approx \arccos\left(1 - \frac{4}{N M}\right) \approx 2\sqrt{\frac{2}{N M}}  .
\end{equation}
Considering the subspace ${\cal I}_-$, there are two eigenvectors which remain relevant in the asymptotic limit  {(for the others the overlap with $\ket{\tau_1}$ vanishes at least as $O(1/\sqrt{NM})$)}, namely
\begin{equation}
\label{diff:evec2} 
\ket{\psi_3^{(\pm)}} = \frac{1}{\sqrt{2}}\ket{\tau_1} \pm \frac{i}{2}(\ket{\tau_9} - \ket{\tau_4}).    
\end{equation}
The eigenvalues are $\lambda_3^{(\pm)} = e^{\pm i\omega_3}$, where $\cos\omega_3$ is the largest root of the quartic equation
\begin{eqnarray}
\nonumber 0 & = & x^4 + \frac{N-2}{d+1} x^3 - \left(1 - \frac{N M}{{(d+1)}^2}\right) x^2 - \\
\nonumber & & - \frac{(N-2)(d-1)}{{(d+1)}^2} x + \frac{N(M-2)}{{(d+1)}^2}.
\end{eqnarray}
Its asymptotic form is given by 
\begin{equation}
\label{different:omega3}
  \omega_3 \approx \arccos\left(1 - \frac{1}{N M}\right) \approx \sqrt{\frac{2}{N M}}.
\end{equation}
From (\ref{diff:evec1}), (\ref{diff:evec2}) we see that for a large graph the sender and the receiver states can be decomposed into the eigenvectors of the evolution operator $\hat U_{s,r}$ according to
\begin{eqnarray}
\nonumber \ket{s} & = & \sqrt{\frac{3}{8}} \ket{\psi_1} + \frac{1}{4} \left(\ket{\psi_2^{(+)}} + \ket{\psi_2^{(-)}}\right) + \\
\nonumber & & + \frac{1}{2} \left(\ket{\psi_3^{(+)}} + \ket{\psi_3^{(-)}}\right) , \\
\nonumber \ket{r} & = & \sqrt{\frac{3}{8}} \ket{\psi_1} + \frac{1}{4} \left(\ket{\psi_2^{(+)}} + \ket{\psi_2^{(-)}}\right) - \\
\nonumber & & - \frac{1}{2} \left(\ket{\psi_3^{(+)}} + \ket{\psi_3^{(-)}}\right) .
\end{eqnarray}
The evolution of STA takes place in a five dimensional subspace
\begin{eqnarray}
\nonumber   \ket{\phi(t)} & = & \sqrt{\frac{3}{8}} \ket{\psi_1} + \\
\nonumber & &  + \frac{1}{4} \left(e^{i \omega_2 t}\ket{\psi_2^{(+)}} + e^{-i \omega_2 t}\ket{\psi_2^{(-)}}\right) + \\
\label{diff:evol} & & + \frac{1}{2} \left(e^{i \omega_3 t}\ket{\psi_3^{(+)}} + e^{-i \omega_3 t}\ket{\psi_3^{(-)}}\right) .
\end{eqnarray}
The fidelity of STA after $t$ steps (\ref{fidelity}) can be expressed as a sum
\begin{equation}
{\cal F}(t) = \sum_{j=5}^8 |\braket{\nu_j}{\phi(t)}|^2 .
\label{fid:diff}
\end{equation}
From (\ref{diff:evec1}), (\ref{diff:evec2}) and (\ref{diff:evol}) we find that the transfer probabilities to individual states $\ket{\nu_j}$ are given by
\begin{eqnarray}
\nonumber |\braket{\nu_5}{\phi(t)}|^2 & = & \frac{1}{64}\left(3 +\cos\left(\omega_2 t\right)  - 4\cos\left(\omega_3 t\right)\right)^2 , \\
\nonumber |\braket{\nu_6}{\phi(t)}|^2 & = & \frac{1}{16} \sin^4\left(\frac{\omega_2 t}{2}\right) , \\
\nonumber |\braket{\nu_7}{\phi(t)}|^2 & = & 0, \\
\label{diff:fid:ind} |\braket{\nu_8}{\phi(t)}|^2 & = & \frac{1}{32}\left(\sin(\omega_2 t) - 2\sin(\omega_3 t)\right)^2 .
\end{eqnarray}
From the asymptotic expansions (\ref{different:omega2}), (\ref{different:omega3}) we see that for a large graph the frequencies are harmonic 
$$
\omega_2 = 2\omega_3.
$$
Hence, we find that with probability 
\begin{equation}
\label{diff:fid:r} |\braket{r}{\phi(t)}|^2 = \sin^8 {\left(\frac{\omega_3 t}{2}\right)},
\end{equation}
the walker is in the receiver state $\ket{r} = \ket{\nu_5}$, i.e., at the receiver vertex in the loop, and with probability
\begin{eqnarray}
\nonumber |\braket{\nu_6}{\phi(t)}|^2 + |\braket{\nu_8}{\phi(t)}|^2 & = & \frac{1}{2} \sin^2\left(\omega_3 t\right)\sin^4\left(\frac{\omega_3 t}{2}\right) + \\
\label{diff:fid:nr} & & + \frac{1}{16}\sin^4\left(\omega_3 t\right) ,
\end{eqnarray}
it is at the receiver vertex but not in the loop. Overall, the fidelity of STA for a large graph is given by
\begin{equation}
    {\cal F}(t) = \sin^4\left(\frac{\omega_3 t}{2}\right) .
\label{diff:fid:overall}
\end{equation}
We conclude that the state transfer is achieved after $T^{(st)}$ steps, where 
\begin{equation}
\label{diff:tmax}
    T^{(st)} \approx \pi\sqrt{\frac{N M}{2}}.
\end{equation}
At this time the walker is with high probability in the receiver state $\ket{r}$.

\begin{figure}
    \centering
    \includegraphics[width=0.45\textwidth]{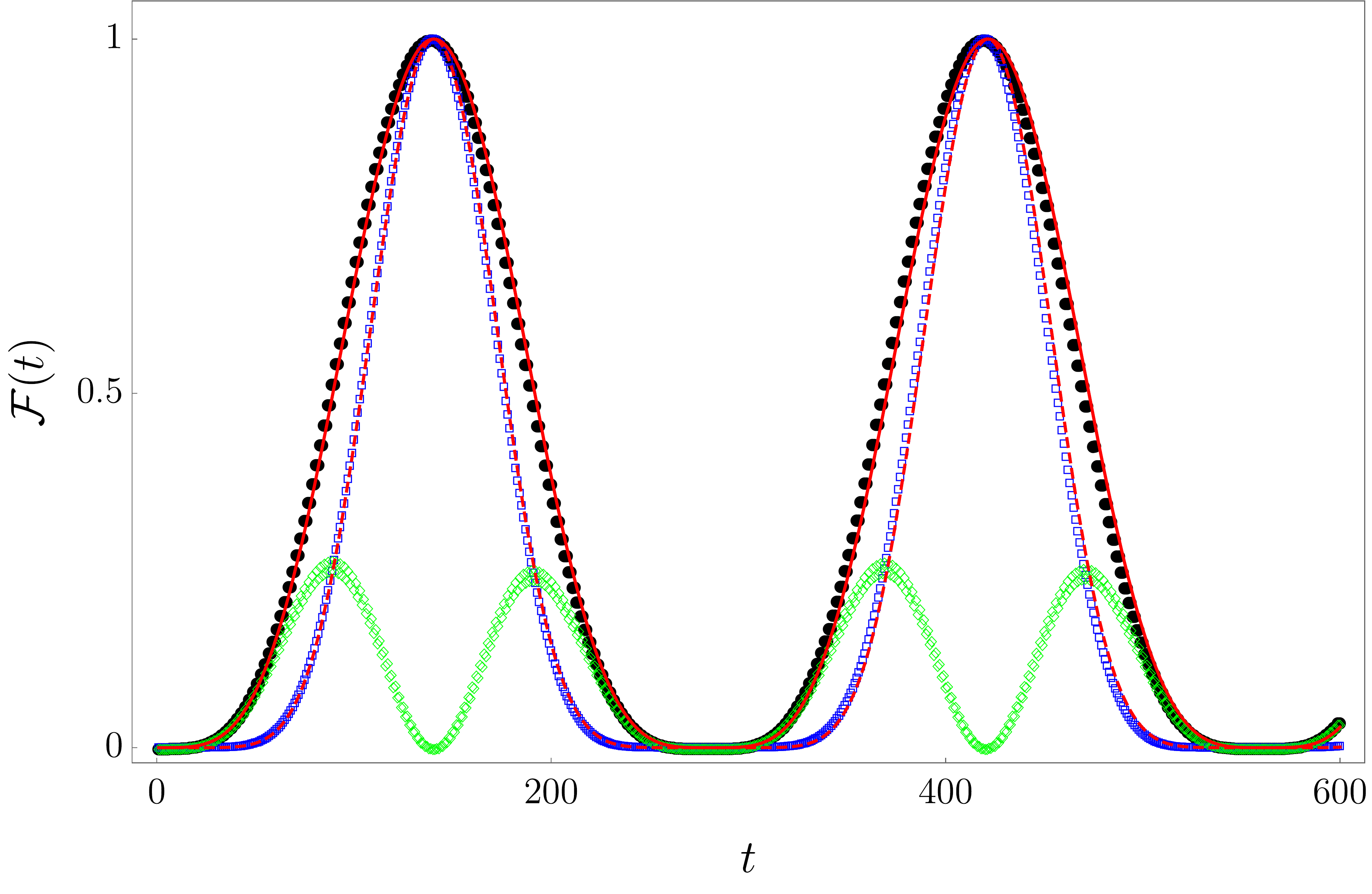}
    \caption{Overall fidelity of the STA as a function of the number of steps $t$ for $N=40$ and $M=100$. The sender and the receiver vertices are in different partitions.  {Black dots are obtained from the numerical simulation, the full red curve corresponds to (\ref{diff:fid:overall}).} The walker is transferred to the receiver vertex with fidelity close to one after $T^{(st)}\approx 140$ steps, in accordance with (\ref{diff:tmax}). At this time the walker is found with high probability in the loop (blue squares), as follows from the analytical prediction (\ref{diff:fid:r})  {depicted by the red dashed curve}. Green diamonds represent the probability that the walker is at the marked vertex but not in the loop, which follows the curve (\ref{diff:fid:nr}).}
    \label{fig:diff}
\end{figure}

For illustration we show in Figure~\ref{fig:diff} the evolution of fidelity for a graph with $N=40$ and $M=100$.

\begin{figure}
    \centering
     \includegraphics[width=0.45\textwidth]{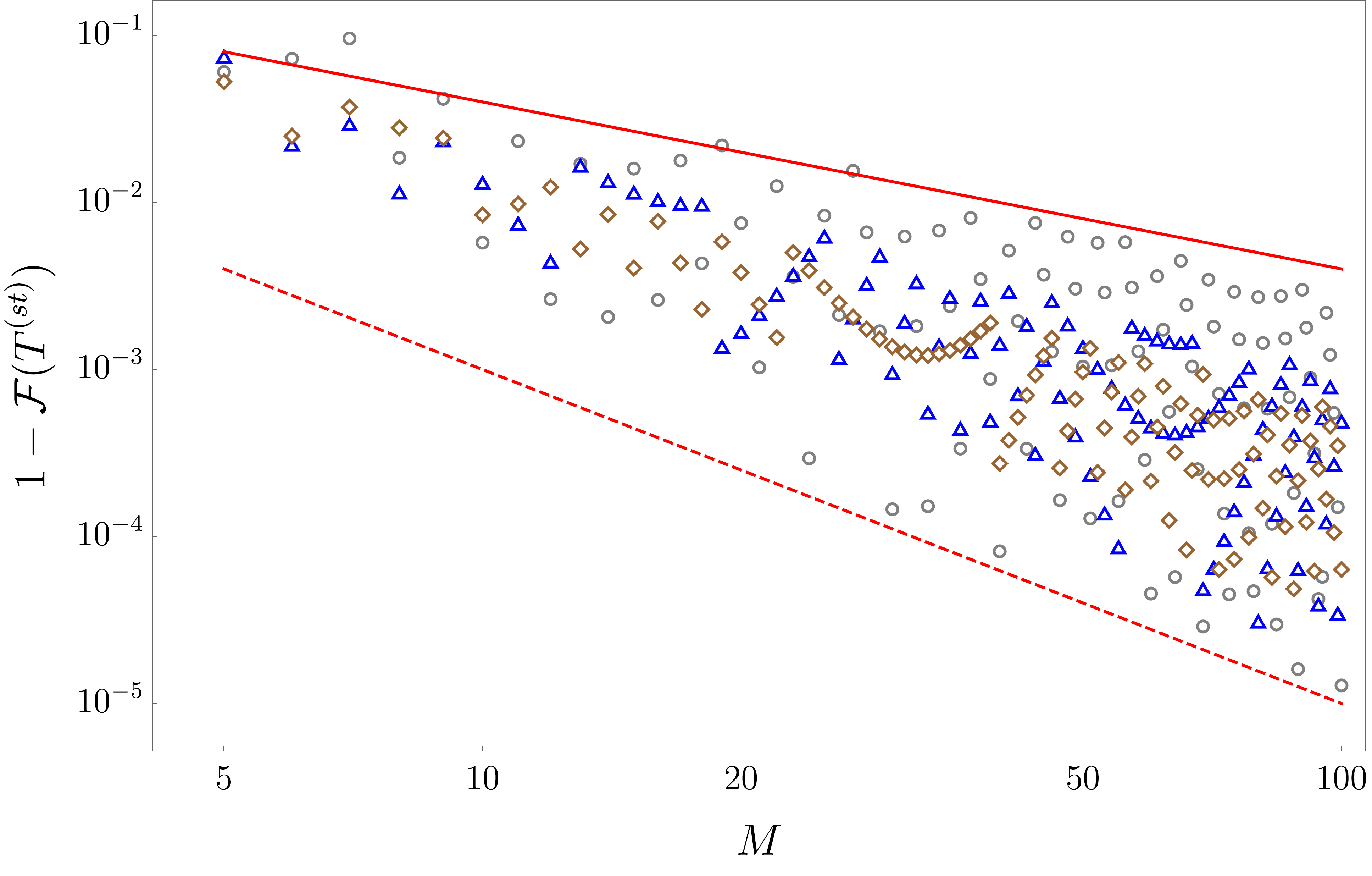}
    \caption{ {Overall fidelity of STA as a function of the number of partitions $M$ for $N=10$ (gray circles), $N=50$ (blue triangles) and $N=100$ (brown diamonds). The sender and the receiver vertices are in different partitions. For a given $N$ and $M$ we evaluate numerically the evolution of STA for the optimal number of steps $T^{(st)}$ given by (\ref{diff:tmax}) and determine ${\cal F}(T^{(st)})$ from the formula (\ref{fid:diff}). To unravel the scaling of the fidelity we plot $1-{\cal F}(T^{(st)})$ on the log-log scale.  The full red line has the $1/M$ slope, while the dashed red line follows $1/M^2$. The plot indicates that $O(1/M)>1-{\cal F}(T^{(st)})>O(1/M^2)$, and that the fluctuations decrease with increasing $N$.}}
    \label{fig:diff:nm}
\end{figure}

 {The result (\ref{diff:fid:overall}) holds in the limit of large $N$ and $M$. To investigate how quickly does the fidelity at the optimal time (\ref{diff:tmax}) approaches unity we performed numerical simulations for various values of $N$ and $M$. The simulations indicate that the fidelity can be again estimated by
$$
{\cal F}(T^{(st)}) = 1 - O\left(\frac{1}{M}\right),
$$
however, the dependence on $N$ is more complex than for search and STA with the sender and the receiver in the same partition. The results are illustrated in Figure~\ref{fig:diff:nm}.
}

\section{State transfer algorithm with an active switch}
\label{sec:5}

As we have shown in the previous section, the STA does not perform with unit fidelity on the complete $M$-partite graph with loops when the sender and the receiver are in the same partition. Note that if the receiver does not know the position of the sender, the measurement should be made at the optimal time (\ref{diff:tmax}) corresponding to the more likely configuration, i.e., when the sender and the receiver are in different partition. This reduces the fidelity of STA further to approximately 0.91.

To fix this issue we introduce an STA where the sender and the receiver will actively switch the local coins at their vertices. We use that for $M\to\infty$ and $N\to\infty$ the state of the SA (\ref{sa:state:t}) on the complete $M$-partite graph with loops evolves periodically from the initial state $\ket{\Omega}$ to the target state $\ket{\nu_1}$ and back with a period of $2T$, where $T$ is the run-time of the SA given by (\ref{search:tmax}). Hence, we can perform the state transfer in the following way. The sender initializes the walk on its vertex in the target state of the search algorithm $\ket{s} = \ket{\nu_1}$ which corresponds to the loop at the sender vertex. For the first $T$ steps only the sender will use the marked coin, i.e., the walk will evolve according to the operator $\hat U_s$ of the SA with the marked vertex $s$. The sender state $\ket{s}$ will evolve close to the equal weight superposition $\ket{\Omega}$, i.e., the initial state of the SA (\ref{InitSearch}). Afterwards, the sender switches off the marked coin, and the receiver switches it on, i.e., the walk evolves according to the operator $\hat U_r$ of the SA with the marked vertex $r$. After another $T$ steps the walk will evolve close to the state $\ket{r}$ corresponding to the loop at the receiver vertex, and the receiver will detect it with high probability. In this way we can achieve state transfer  {with high fidelity} on the complete $M$-partite graph with loops irrespective of the relative position of the sender and the receiver. 

\begin{figure}
    \centering
    \includegraphics[width=0.45\textwidth]{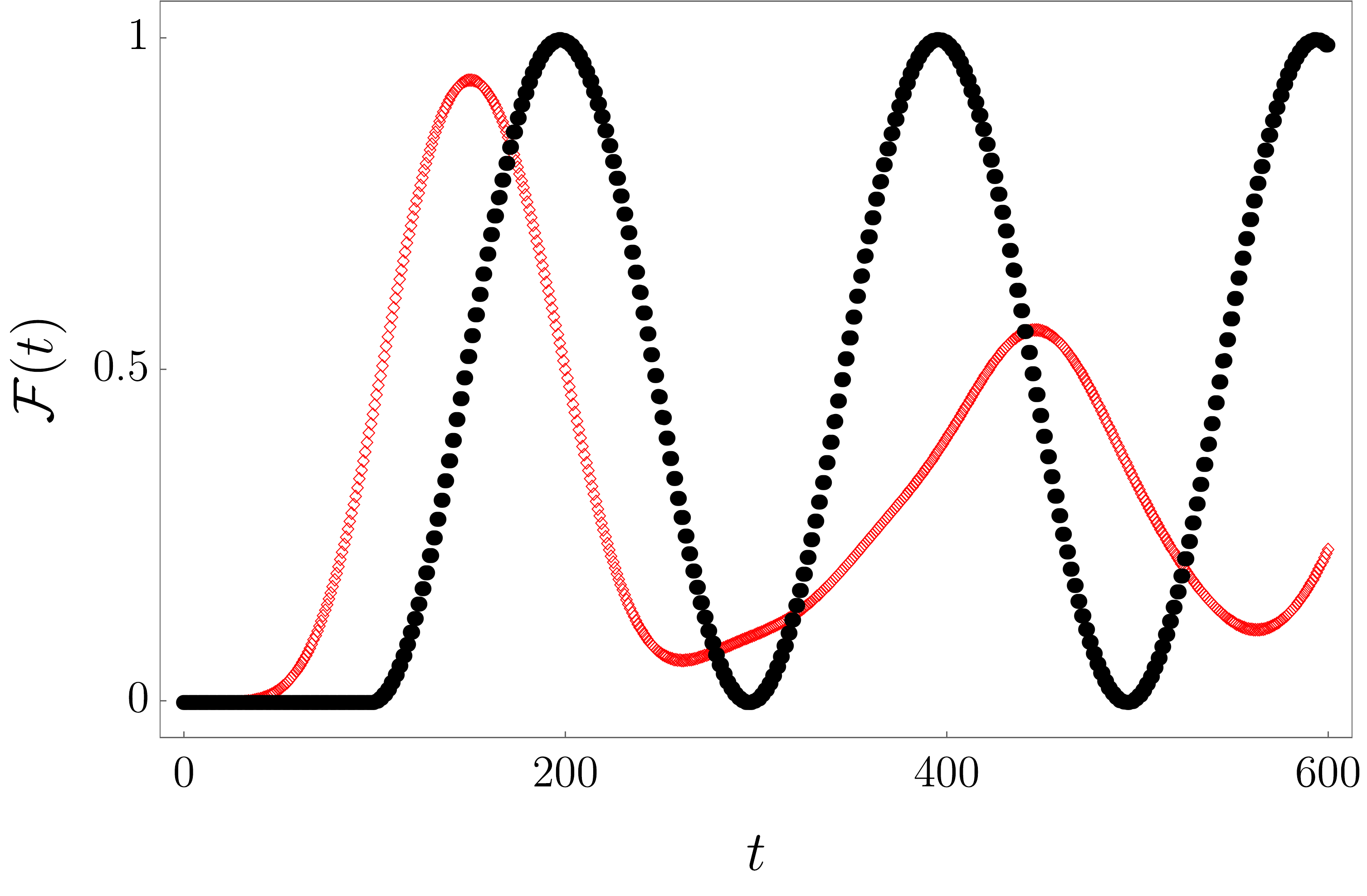}
    \caption{Comparison of fidelity of the original STA (\ref{same:fid}) (red diamonds) with fidelity of the STA with an active switch (black dots) as a function of the number of steps $t$ for $N=40$ and $M=100$. The sender and the receiver vertices are in the same partition. The switch between $\hat{U}_s$ and $\hat{U}_r$ is done after $T \approx 100$ steps, corresponding to the run-time of the SA.}
    \label{fig:comp}
\end{figure}

For illustration, we show in Figure~\ref{fig:comp} the comparison of the fidelities of state transfer of the original STA and the STA with an active switch when the sender and the receiver are in the same partition. We see that the STA with an active switch takes more steps, however, the fidelity reaches one.

Let us now formalize the STA with an active switch on more general graphs. Namely, we consider graphs where the optimal number of steps $T$ of the SA does not depend on the position of the marked vertex $m$. The steps of the STA with an active switch can be formulated as follows:
\begin{enumerate}
\item Sender initializes the walk at its vertex in the state $\ket{s}$ corresponding to the target state of the SA with the marked vertex $s$.  
\item  {Sender uses marked coin on his vertex for $T$ steps, i.e.,} the evolution operator $\hat{U}_{s}$ is applied $T$-times.
\item  {Receiver uses marked coin on his vertex for $T$ steps, i.e.,} the evolution operator $\hat{U}_{r}$ is applied $T$-times.
\item Receiver measures the walk at its vertex.
\end{enumerate}

We show that the fidelity of the STA with an active switch can be lower bounded using only the results from the SA with one marked vertex, which is not true for the original STA. As we have seen in the case of the complete $M$-partite graph with loops, the SA does not tell us anything about the evolution of the STA in the subspace ${\cal I}_{-}$.

To derive the lower bound of fidelity of the STA with an active switch we first introduce two conditions on the SA. The first condition is related to target state of the SA. We suppose that after $T$ steps the state of the SA can be expressed in the form
\begin{eqnarray}
\ket{\phi(T)} = \hat{U}_m^{T}\ket{\Omega} = \alpha_m\ket{m} + \epsilon_m\ket{\eta_m}
\label{firstCondit}
\end{eqnarray}
for every marked vertex $m$ in the graph. Here $\ket{m}\in {\cal H}_m$ is the target state of the SA, i.e., if the walk is in this state the success probability of finding the marked vertex $m$ is exactly $1$, and $\ket{\eta_m}$ is a unit vector orthogonal to $\ket{m}$. Complex numbers $\alpha_m$ and $\epsilon_m$ are such that $\vert\alpha_m\vert$ is close to one and $\vert\epsilon_m\vert \ll 1$.
$\vert\alpha_m\vert^2$ is closely related to the success probability of the SA, since
\begin{eqnarray}
\nonumber P_m(T) & = &\vert\braket{m}{\phi(T)}\vert^2+\sum_{\substack{\ket{j}\in {\cal H}_m\\ \braket{j}{m}=0}}\vert\braket{j}{\phi(T)}\vert^2 \\ & = & \vert\alpha_m\vert^2+\vert\epsilon_m\vert^2\sum_{\substack{\ket{j}\in {\cal H}_m\\ \braket{j}{m}=0}}\vert\braket{j}{\eta_m}\vert^2 .
\end{eqnarray}
Note that if the vector $\ket{\eta_m}$ does not have a support at the marked vertex then the success probability is exactly $\vert\alpha_m\vert^2$. From the relation (\ref{firstCondit}) for $m=s$ we express the initial state of the STA with an active switch as
\begin{equation}
    \ket{s} = \frac{1}{\alpha_s}\left( \hat U_s^T\ket{\Omega} - \epsilon_s\ket{\eta_s}\right) .
\label{staas:s}
\end{equation}
The second condition describes the periodicity of the SA. Namely, we suppose that after $2T$ steps the state of the SA can be written in the form
\begin{eqnarray}
\hat{U}_m^{2T}\ket{\Omega} = \beta_m\ket{\Omega} + \delta_m\ket{\rho_m} ,
\label{secondCondit}
\end{eqnarray}
where $\ket{\rho_m}$ is a unit vector orthogonal to the initial state $\ket{\Omega}$. $\beta_m$ and $\delta_m$ are again complex numbers where $\vert\beta_m\vert^2$ is the return probability. We assume that it is close to one and that $\vert\delta_m\vert \ll 1$. In other words, this condition says that if we double the number of steps the SA returns close to its initial state.

Assuming that the SA satisfies the conditions (\ref{firstCondit}) and (\ref{secondCondit}) we write the final state of the STA with an active switch in the following manner 
\begin{eqnarray}
\nonumber \hat{U}_r^{T}\hat{U}_s^{T}\vert s\rangle & = & \frac{1}{\alpha_s}\hat{U}_r^{T}\hat{U}_s^{T}\left(\hat{U}_s^{T}\vert \Omega\rangle-\epsilon_s\vert \eta_s\rangle\right) \\ 
\nonumber & = & \frac{1}{\alpha_s}\hat{U}_r^{T}\left(\hat{U}_s^{2T}\vert \Omega\rangle-\epsilon_s\hat{U}_s^{T}\vert \eta_s\rangle\right) \\
\nonumber & = & \frac{1}{\alpha_s}\hat{U}_r^{T}\left(\beta_s\vert \Omega\rangle+\delta_s\vert \rho_s\rangle-\epsilon_s\hat{U}_s^{T}\vert \eta_s\rangle\right) \\
\nonumber  & = & \frac{\alpha_r}{\alpha_s}\beta_s\vert r\rangle+\frac{\beta_s}{\alpha_s}\epsilon_r\vert \eta_r\rangle+\frac{\delta_s}{\alpha_s}\hat{U}_r^{T}\vert \rho_s\rangle- \\
& & - \frac{\epsilon_s}{\alpha_s}\hat{U}_r^{T}\hat{U}_s^{T}\vert \eta_s\rangle
\label{calculFinalState}
\end{eqnarray}
where we have first used (\ref{staas:s}), then (\ref{secondCondit}) and finally we again use (\ref{staas:s}) but for the state $\vert r\rangle$. The fidelity of STA with an active switch can be expressed as
$$
{\cal F} = \left\vert\left\langle r\left\vert\hat{U}_r^{T}\hat{U}_s^{T}\right\vert s\right\rangle\right\vert^2 + \sum_{\substack{\ket{j}\in {\cal H}_r\\ \braket{j}{r}=0}}\left\vert\left\langle j\left\vert\hat{U}_r^{T}\hat{U}_s^{T}\right\vert s\right\rangle\right\vert^2 .
$$
Hence, the square root of the fidelity can be bounded from below by
$$
\sqrt{\cal F} \geq \left\vert\left\langle r\left\vert\hat{U}_r^{T}\hat{U}_s^{T}\right\vert s\right\rangle\right\vert .
$$
To approximate $\left\vert\left\langle r\left\vert\hat{U}_r^{T}\hat{U}_s^{T}\right\vert s\right\rangle\right\vert$ we use the following estimates 
\begin{eqnarray}
\nonumber \left\vert\left\langle r\left\vert\hat{U}_r^{T}\right\vert \rho_s\right\rangle\right\vert & \leq & \vert\vert\vert r\rangle\vert\vert \left\vert\left\vert\hat{U}_r^{T}\left\vert \rho_s\right\rangle\right\vert\right\vert=\vert\vert\vert r\rangle\vert\vert \left\vert\left\vert\left\vert \rho_s\right\rangle\right\vert\right\vert = 1, \\
\left\vert\left\langle r\left\vert\hat{U}_r^{T}\hat{U}_s^{T}\right\vert \eta_s\right\rangle\right\vert  & \leq &  \vert\vert\vert r\rangle\vert\vert \left\vert\left\vert\hat{U}_r^{T}\hat{U}_s^{T}\left\vert \eta_s\right\rangle\right\vert\right\vert=1 ,
\label{estimation}
\end{eqnarray}
which follow from the Cauchy–Schwarz inequality and the unitarity of evolution operator $\hat U_m$. Combining (\ref{calculFinalState}) with (\ref{estimation}) we find the lower bound for the square root of the fidelity which reads
\begin{widetext}
\begin{eqnarray}
\nonumber \sqrt{\cal F}  & \geq & \left\vert \frac{\alpha_r}{\alpha_s}\beta_s\braket{r}{r}+\frac{\beta_s}{\alpha_s}\epsilon_r\braket{r}{\eta_r}+\frac{\delta_s}{\alpha_s}\left\langle r\left\vert\hat{U}_r^{T}\right\vert \rho_s\right\rangle-\frac{\epsilon_s}{\alpha_s}\left\langle r\left\vert\hat{U}_r^{T}\hat{U}_s^{T}\right\vert \eta_s\right\rangle\right\vert \\
\nonumber & \geq &\frac{\vert\alpha_r\vert}{\vert\alpha_s\vert}\vert\beta_s\vert-\frac{\vert\delta_s\vert}{\vert\alpha_s\vert}\left\vert\left\langle r\left\vert\hat{U}_r^{T}\right\vert \rho_s\right\rangle\right\vert-\frac{\vert\epsilon_s\vert}{\vert\alpha_s\vert}\left\vert\left\langle r\left\vert\hat{U}_r^{T}\hat{U}_s^{T}\right\vert \eta_s\right\rangle\right\vert  \\
& \geq & \frac{\vert\alpha_r\vert}{\vert\alpha_s\vert}\vert\beta_s\vert-\frac{\vert\delta_s\vert}{\vert\alpha_s\vert}-\frac{\vert\epsilon_s\vert}{\vert\alpha_s\vert} 
\label{lowerBoundary}.
\end{eqnarray}
\end{widetext}
It is easy to see from (\ref{lowerBoundary}) that if $\vert\alpha_s\vert$, $\vert\alpha_r\vert$ and $\vert\beta_s\vert$ are close to one and if $\vert\epsilon_s\vert\ll1$ and $\vert\delta_s\vert\ll1$ then the fidelity of the state transfer is close to one. 

 {The result derived above guarantees that if the SA succeeds with unit probability in the limit of a large graph, then the STA with an active switch achieves perfect state transfer. Moreover, even for small graphs the STA with an active switch can actually achieve very good fidelity. For illustration we have investigated numerically the STA with an active switch on the complete $M$-partite graph for various values of $N$ and $M$. The results are similar to those presented in Figure~\ref{fig:diff:nm}.}

\begin{figure}[ht]
 \centering
 \includegraphics[width=0.45\textwidth]{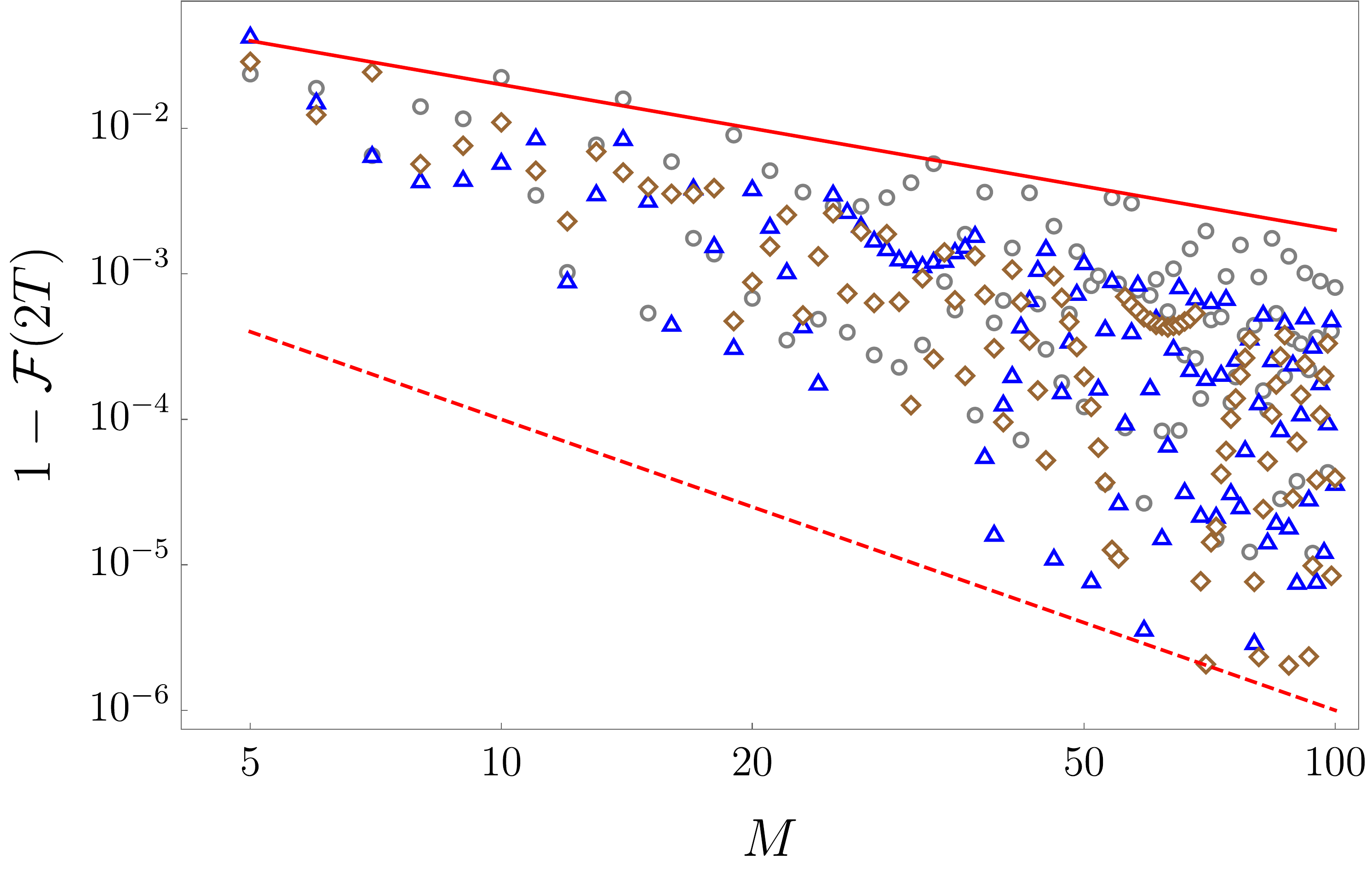}
 \caption{ {Overall fidelity of STA with an active switch as a function of the number of partitions $M$ for $N=10$ (gray circles), $N=50$ (blue triangles) and $N=100$ (brown diamonds). The walk is initialized at the sender vertex in the loop. For a given $N$ and $M$ we evaluate numerically the evolution operator $\hat U_s$ of SA with marked vertex $s$, apply it for the optimal number of steps $T$ given by (\ref{search:tmax}). Then we repeat the same with $\hat U_r$. Finally, we make a measurement at the receiver vertex $r$ and determine the fidelity ${\cal F}(2T)$. To unravel the scaling of the fidelity we plot $1-{\cal F}(2T)$ on the log-log scale. The full red line has the $1/M$ slope, while the dashed red line follows $1/M^2$. The plot indicates that $O(1/M)>1-{\cal F}(2T)>O(1/M^2)$.}}
 \label{fig:9}
\end{figure}

\section{Conclusion}
\label{sec:6}

We have investigated search and state transfer algorithms based on the coined quantum walks, focusing on the complete $M$-partite graph. It was shown that adding loops to all vertices increases the success probability of the SA close to one for a large graph. This is by now a standard method \cite{Ambainis2005,Potocek2009,wong2015,wong2018,rhodes2019,rhodes2020,chiang2020} which has a potential to significantly improve the success probability on a much broader class of graphs \cite{hoyer2020}. However, the analysis of the SA does not provide the necessary insight for the investigation of the STA, as the latter requires larger invariant subspace. As we have seen on the example of the complete $M$-partite graph with loops, success probability of the SA close to one does not guarantee STA with unit fidelity. The reason is that the phases of the relevant eigenvalues of the evolution operator are not harmonic when the sender and the receiver are in the same partition.  {Although the modification of the initial state has improved the fidelity considerably, the absence of an edge between the sender and the receiver vertex on the complete $M$-partite graph does not allow for perfect state transfer.} It would be interesting to find out if this occurs for different graphs as well. 

In the present paper we have limited our investigations to the case where all partitions have the same number of vertices $N$. This enabled us to find exact invariant subspaces for SA and STA which have dimensions independent of $N$ and $M$. Allowing the partitions with different number of vertices appears to break this feature, and the dimension of the invariant subspace is likely to depend on $M$. We plan to investigate this behaviour in the near future for small values of $M$.

To improve the STA on the complete $M$-partite graph we have introduced the STA with an active switch. This approach allows for perfect state transfer  {in the limit of a large graph} in both configurations of the sender and the receiver vertex. The trade-off is that the STA with an active switch requires more steps than the original STA. Indeed, when the sender and the receiver are in different partitions, the number of steps for the original STA to reach unit fidelity is twice the number of steps of the search for two vertices. On the other hand, STA with an active switch takes twice the number of steps of the SA for one vertex. Since the search for two vertices is $\sqrt{2}$ faster than the search for one vertex, the STA with an active switch is slower by the same factor.

The main advantages of the STA with an active switch are that it can be applied to other graphs, and that its fidelity can be estimated based on the analysis of the SA for one marked vertex alone. In this way we can achieve state transfer with high fidelity on graphs, where the SA for one marked vertex has success probability close to one and evolves almost periodically. For many symmetric graphs these conditions are well satisfied, at least in the limit of a large graph. Moreover, exact periodicity of the Grover walk (i.e., without the marked vertex) was recently investigated for various graphs \cite{kubota2018,kubota2019,yoshie2019}. It would be of interest to determine if SA works on these graphs as well.

The STA with an active switch also has some disadvantages. As we have already mentioned, it will have a longer run-time in comparison with the original STA. Moreover, the sender and the receiver have to actively switch off or on their marked coins. However, this is only a local operations, and since we consider that the run-time of the SA is independent of the location of the marked vertex, the time of the switching is determined solely by the global properties of the graph, e.g., the number of vertices. Hence, the sender and the receiver still do not need to know each other's position. Finally, we have to determine the target state of SA, which serves as the initial state for the STA with an active switch. Nevertheless, for highly symmetric graphs this target state is usually either the equal weight superposition of all direction or the state corresponding to a loop.

\begin{acknowledgments}

Both authors received support from the Czech Grant Agency (GA\v CR) under grant No.~17-00844S and from MSMT RVO 14000. This publication was funded by the project ``Centre for Advanced Applied Sciences", Registry No.~CZ.$02.1.01/0.0/0.0/16\_019/0000778$, supported by the Operational Programme Research, Development and Education, co-financed by the European Structural and Investment Funds and the state budget of the Czech Republic. 
\end{acknowledgments}


\bibliographystyle{apsrev4-2}
\bibliography{biblio}

\end{document}